\documentclass[aps, superscriptaddress, floatfix, 12pt]{revtex4}
\usepackage{graphicx}
\usepackage{amsmath}
\usepackage{amssymb}

\newcommand{\be}{\begin{eqnarray}}
\newcommand{\ee}{\end{eqnarray}}

\newcommand{\el}{\nonumber \hfill \\}
\newcommand{\Tr}{\mathrm{Tr}}
\newcommand{\nn}{\nonumber }
\def\slash#1{\setbox0=\hbox{$#1$}               
   \dimen0=\wd0                                 
   \setbox1=\hbox{/} \dimen1=\wd1               
   \ifdim\dimen0>\dimen1                        
      \rlap{\hbox to \dimen0{\hfil/\hfil}}      
      #1                                        
   \else                                        
      \rlap{\hbox to \dimen1{\hfil$#1$\hfil}}   
      /                                         
   \fi}                                         %

\begin{document}

\title{Influence of Quark Boundary Conditions on the Pion Mass in
  Finite Volume}
\author{J. Braun}
\affiliation{Institute for Theoretical Physics, University of
  Heidelberg, Philosophenweg 19, 69120 Heidelberg}
\author {B. Klein} 
\affiliation{GSI, Planckstrasse 1, 64159 Darmstadt}
\author{H.J. Pirner}
\affiliation{Institute for Theoretical Physics, University of
  Heidelberg, Philosophenweg 19, 69120 Heidelberg}
\affiliation{Max-Planck-Institut f\"ur Kernphysik, Saupfercheckweg 1,
  69117 Heidelberg}

\date{\today}

\begin{abstract}

We calculate the mass shift for the pion in a finite volume
with renormalization group (RG) methods in the framework of the
quark-meson model.
In particular, we investigate the importance of the quark effects on
the pion mass. As in lattice gauge theory, the choice of quark
boundary conditions has
a noticeable effect on the pion 
mass shift in small volumes, in addition to the shift due to
pion interactions. We compare our results to chiral perturbation
theory calculations and  
find differences due to the fact that chiral perturbation theory only 
considers pion effects in the finite volume.

\end{abstract}

\maketitle

\section{Introduction}

In the study of QCD, non-perturbative methods are essential in order to
understand the connection between the high-momentum regime dominated
by quarks and gluons, and the low-momentum regime described in terms
of hadronic degrees of freedom. Lattice gauge theory is
a method of great importance in this quest. Current
simulations with dynamical fermions are
limited to rather small lattice sizes and in some approaches to
quark masses which are still large compared to the physical values. 
In addition to taking the continuum
limit in which the lattice spacing is taken to zero, results from
lattice calculations require extrapolation towards the chiral limit
and the thermodynamic limit. Thus, in order to compare a result for an
observable simulated in a
small volume with the physical observable, it is
essential to understand the finite volume effects.
Apart from the application to lattice QCD, these finite volume effects are
also interesting in their own right and worth investigating.

The most important tool for extrapolations of lattice gauge theory
results to small pion masses and to large volumes is chiral
perturbation theory (chPT) \cite{Gasser:1986vb, Gasser:1987zq,
  Procura:2003ig, Bernard:2003rp, 
  Bernard:2005fy, Arndt:2004bg,
  AliKhan:2003cu, Colangelo:2003hf, Colangelo:2004xr,
  Colangelo:2005gd, Bedaque:2004dt}. 
In particular for the chiral extrapolation to small pion masses
\cite{Leinweber:1998ej, Detmold:2001jb, Procura:2003ig, Bernard:2003rp,
  Bernard:2005fy}, and for the
extrapolation to infinite volume for
properties of the nucleon \cite{AliKhan:2003cu}, chiral perturbation
theory describes the lattice results very well.

In contrast to these applications, the finite volume 
shifts of the meson masses are less well described by chiral
perturbation theory \cite{Aoki:2002uc, Guagnelli:2004ww, Orth:2005kq}.
For the pion mass, the shifts predicted by chiral perturbation theory are
consistently smaller than those observed in lattice simulations.
We expect that chiral perturbation theory correctly
  describes finite volume effects for volumes that are suffciently
  large so that the internal degrees of freedom
  such as quarks and gluons are unimportant \cite{Colangelo:2005gd}. 
The discrepancies in current systematic investigations of finite
volume effects \cite{AliKhan:2003cu, Aoki:2002uc, Guagnelli:2004ww,
  Orth:2005kq, AliKhan:2001tx} 
seem to indicate that there this range has not yet been reached.

\begin{figure}
\includegraphics[scale=0.9, clip=true, angle=0,
  draft=false]{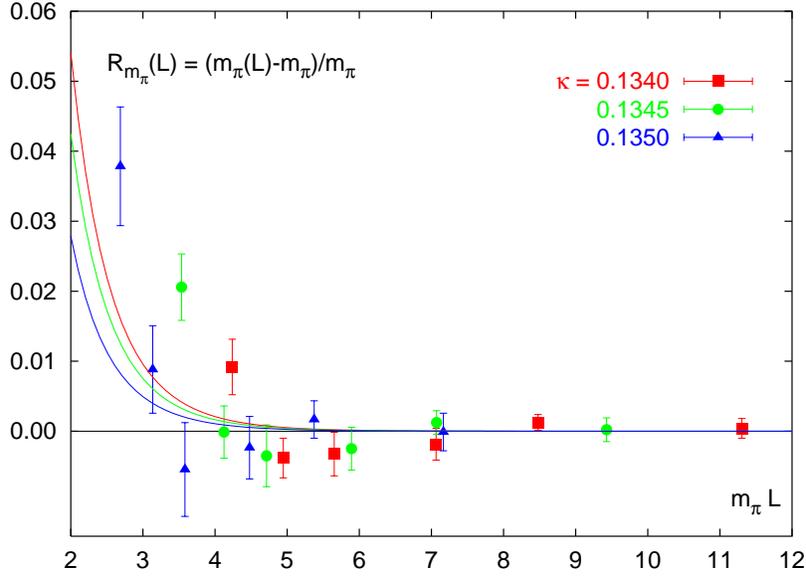} 
\caption{\label{fig:zero} The pion mass shift $R[m_\pi(L)] =
(m_\pi(L)-m_\pi(\infty))/m_\pi(\infty)$ as a function of
  $m_\pi(\infty) \cdot L$, obtained in a quenched lattice
  calculation, from ref.~\cite{Guagnelli:2004ww}. Shown are results 
  for three different 
  values of the quark mass, determined by $\kappa$. The solid lines
  show the corresponding 
  predictions from chiral perturbation theory.\\}
\includegraphics[scale=0.45, clip=true, angle=0,
  draft=false]{fig2.eps} 
\caption{\label{fig:orth} The pion mass $m_\pi(L)$ as
  a function of box size 
  $L$, obtained in a lattice simulation with two flavors of dynamical
  Wilson fermions, from ref.~\cite{Orth:2005kq}. Results for pion
  masses $m_\pi=643 \;\mathrm{MeV}, 490\; \mathrm{MeV}, 419 \;
  \mathrm{MeV}$ (circles, squares, diamonds) are compared to results
  from L\"uscher's formula 
  \cite{Luscher:1985dn}  with input from chPT up to NNLO order
  \cite{Colangelo:2003hf}. For details, see
  ref.~\cite{Orth:2005kq}. For the smallest pion mass, a drop similar
  to the one in Fig.~\ref{fig:zero} can be observed.}
\end{figure}

One issue which cannot be addressed by chiral perturbation theory
alone is the
influence of the boundary conditions for the quark fields. 
While fermionic fields require anti-periodic
boundary conditions in the Euclidean time direction, we are free to
choose either periodic (p. b.c.) or anti-periodic (a.p. b.c.) boundary
conditions in the spatial directions. 
In lattice calculations, 
this choice changes the finite size effects: 
In the investigation \cite{Aoki:1993gi}
of finite volume effects it was
found that the choice of the boundary conditions for the quark fields
has a direct influence on the size of the observed finite volume
shifts \cite{Aoki:1993gi, Aoki:1993fq} 
and an explanation in terms of quark effects was proposed,
  for both quenched and unquenched calculations.
Such effects cannot be captured by a description in terms of pion fields only.
It has been shown by Gasser and Leutwyler 
  that the low-energy constants in the chiral 
  perturbation theory Lagrangian remain unchanged from their values in infinite
  volume if one considers QCD in a finite Euclidean volume, provided the same
  anti-periodic boundary conditions as in the
  temporal direction are chosen as well in the spatial directions for the
  quark fields \cite{Gasser:1987zq}.
This leaves open the possibility that a change of boundary
  conditions for the quark fields might in fact lead to a change
  in the finite volume behavior.

To motivate further our interest in the influence of boundary conditions, in  
Fig.~\ref{fig:zero}, we present 
an example of a lattice calculation in the
quenched approximation from the ZeRo-collaboration \cite{Guagnelli:2004ww}.
Shown is the shift of the pion mass  $m_\pi(L)$ in finite volume
relative to the value in infinite volume $m_\pi(\infty)$ as a
function of $m_\pi(\infty) \cdot L$ where $L$ is the volume size. Surprisingly,
these results show a  
dropping pion mass for intermediate volume sizes in a region where
the standard chiral perturbation theory result (indicated by the solid
lines in the figure) predicts only a very weak volume dependence. This
behavior  
would be unexpected from pion effects alone.
In addition, finite volume effects from chiral perturbation theory are
predicated on the presence of a ``pion cloud'', which in turn requires
the presence of sea quarks \cite{AliKhan:2003cu}.
Like chPT, the present work is not directly concerned
  with the quenched 
  approximation, which requires its own low-energy effective theory
  \cite{Sharpe:1992ft, Bernard:1992mk}. 
Although the quenched calculation in Fig.~\ref{fig:zero}
  shows the pion mass drop in a very
  distinct fashion, similar effects for the meson
masses are also seen in studies of finite size effects with dynamical quarks
\cite{Aoki:2002uc, Orth:2005kq}. In Fig.~\ref{fig:orth}, we show
  results for the pion mass as a function of the volume size from a
  lattice calculation with two dynamical flavors of Wilson fermions
  \cite{Orth:2005kq}.  Results are given only for few volume
  sizes, but they also show a decrease of the pseudoscalar mass in
  small volumes.
 
In this paper, we investigate the volume dependence of the
pion mass in the quark meson model using renormalization group (RG) methods.
In particular, our purpose is to extend our previous work
\cite{Braun:2004yk} and to investigate the influence of
different boundary conditions for the fermionic fields on the finite
volume effects for low-energy observables such as the pion mass and
the pion decay constant. Since our model includes dynamical meson
fields, and a dynamical breaking of the chiral flavor symmetry
according to $SU(N_f)\times SU(N_f) \rightarrow SU(N_f)$ for $N_f=2$
flavors of quarks, our results are applicable to
unquenched lattice calculations with two dynamical quark flavors.

The quark-meson model cannot predict
the volume dependence of pion mass and pion decay constant exactly. 
It is not a gauge theory and thus has neither  
gluons nor quark confinement. At moderate energies, below the
hadronic mass scale, unconfined 
constituent quarks appear instead of baryonic degrees of freedom. On
the other hand, the model 
has been rather successfully employed in the description of the chiral phase
transition \cite{Jungnickel:1995fp, Schaefer:1999em,
  Schaefer:2004en}. The low-energy couplings of the linear sigma-model
with quarks are compatible with those of chiral perturbation theory
\cite{Jungnickel:1997yu}. As we have previously shown \cite{Braun:2004yk}, for
small pion masses and large volumes our results for the volume dependence agree
with those of chiral perturbation 
theory \cite{Colangelo:2003hf, Colangelo:2004xr}, if we apply
anti-periodic boundary conditions for 
the quark fields in the spatial directions. In this paper, we investigate
the effect 
of different boundary conditions for the quark fields on
low-energy observables, namely the pion mass and pion decay constant, in more
detail. 

In spite of the shortcomings of our model, we believe that the current approach
can shed 
light on lattice results regarding the volume dependence of the pion
mass \cite{Aoki:1993gi, Guagnelli:2004ww, Orth:2005kq}. While the
actual mechanism in QCD may be different due to the
presence of color interactions, the approach employed in the present paper
gives a possible explanation for the apparent drop in the pseudoscalar
pion mass in small
volumes observed in \cite{Aoki:2002uc, Guagnelli:2004ww, Orth:2005kq},
which precedes the rise of this mass due to chiral symmetry
restoration in extremely small volumes. 

The paper is organized as follows: In section~\ref{sec:setup}, we
briefly introduce the model and the renormalization group equations
which govern the RG flow in a finite volume. In
section~\ref{sec:calculation},  we solve the flow equations
numerically and in
section~\ref{sec:results} we present the results for the
volume dependence of the pion
mass. A comparison to lattice results and our conclusions are found in
section~\ref{sec:conclusion}. 

\section{RG-Flow equations for the Quark-Meson model}
\label{sec:setup}

As motivated in the introduction, we will use an $O(4)$-invariant linear
$\sigma$-model with $N_f^2=4$ mesonic degrees of freedom $(\sigma,
\vec{\pi})$, coupled to $N_f=2$ flavors of constituent quarks in
an $SU(2)_{L}\otimes SU(2)_{R}$ invariant way.  
This model does not contain gluonic degrees of freedom, and it is not
confining, but it is an effective low-energy model for dynamical
spontaneous chiral symmetry breaking at intermediate scales of $k \lesssim
\Lambda_{UV}$. 
The ultraviolet scale $\Lambda_{UV} \approx 1.5\;\mathrm{GeV}$ 
is determined by the validity of a hadronic representation of QCD. 
At the UV scale $\Lambda_{UV}$, the quark-meson-model is defined by
the bare effective action
\be
\Gamma_{\Lambda_{UV}}[\phi]&=& \int d^4 x
\left\{ \bar{q}( \slash{\partial}+ g m_c)q +
g\bar{q}(\sigma+i\vec{\tau}\cdot\vec{\pi}
\gamma_{5}) q +
\tfrac{1}{2}(\partial_{\mu}\phi)^{2}+U_{\Lambda_{UV}}(\phi)  
\right\} 
\ee
with a current quark mass term $g m_c$ which explicitly breaks the
chiral symmetry. 
The mesonic potential is characterized by
two couplings: 
\be
U_{\Lambda_{UV}}(\phi)=\frac{1}{2}m_{UV}^2\phi^2 +
\frac{1}{4}\lambda_{UV}(\phi^2)^2\,.
\label{eq:pot_UV}
\ee
In a Gaussian approximation, we obtain the one-loop effective
action for the scalar fields $\phi$, 
\be
\label{eq:1loop}
\Gamma[\phi]&=& \Gamma_{\Lambda_{UV}}[\phi] -
\Tr\log\left(\Gamma_{F}^{(2)}[\phi]\right) +
\frac{1}{2}\Tr\log\left(\Gamma_{B}^{(2)}[\phi]\right) 
\ee
where $\Gamma_{B}^{(2)}[\phi]$ and $\Gamma_{F}^{(2)}[\phi]$ are the
inverse two-point functions for the bosonic and fermionic fields,
evaluated at the vacuum expectation value of the mesonic
field $\phi$. We consider the effective action $\Gamma$ in a local
potential approximation (LPA), where the expectation value is taken to be
constant over 
the entire
volume. In order to regularize the functional traces, we use the Schwinger
proper time representation of the logarithms.
A scale dependence is introduced through an infrared cutoff function 
\be
k\frac{\partial}{\partial k} f_{a}(\tau k^{2}) & = &
-\frac{2}{\Gamma(a+1)}(\tau k^{2})^{a+1}e^{-\tau
k^{2}}\, ,\label{eq:cutoff-fct}
\ee
which regularizes the Schwinger proper time integral. By replacing the bare
coupling in the inverse two-point functions with the scale-dependent
running couplings, we obtain a renormalization group flow equation for the
effective potential in infinite volume for zero temperature:
\be
\label{eq:IV_fe}
k \frac{\partial}{\partial k}U_k(\sigma,\vec{\pi}^{2},T\rightarrow\infty,L \to
\infty)={\displaystyle \frac{k^{2(a+1)}}{16a(a-1)\pi^2} \Bigg\{ -\frac{4 N_c
N_f}{(k^2+ M_q^2(\sigma,\vec{\pi}^{2}))^{a-1}} } \nn  \\
 {\displaystyle + \frac{1}{(k^2 + M_{\sigma}^2(\sigma,\vec{\pi}^{2})^{a-1}}
    + \frac{N_f^2-1}{(k^2 + M_{\pi}^2(\sigma,\vec{\pi}^{2}))^{a-1}} \Bigg\} }
\end{eqnarray}
In infinite volume, $a=2$ is the lowest possible integer value we can choose
in order to be able to
perform the Schwinger-proper time integration. Note that in LPA, the effective
action reduces to the effective potential through the relation
\be
\Gamma_k[\phi] &=& \int d^4 x \; U_k(\sigma,\vec{\pi}^{2}).
\ee

Due to the fact that we allow for explicit symmetry breaking, the effective
potential becomes a function of $\sigma$ and $\vec{\pi}^2$. The meson
masses are 
the eigenvalues of the second
derivative matrix of the mesonic potential: 
\be
M_1^2&=& \frac{1}{2} \left[ 2\, U_{\vec{\pi}^2} + 4 \, \vec{\pi}^2 \,
  U_{\vec{\pi}^2 \vec{\pi}^2} + U_{\sigma \sigma} +
  \sqrt{(2\, U_{\vec{\pi}^2} + 4\, \vec{\pi}^2
      U_{\vec{\pi}^2 \vec{\pi}^2} - U_{\sigma \sigma} )^2 
          + 16\, \vec{\pi}^2\, U_{\sigma \vec{\pi}^2}^2
   } \right], 
\el
M_2^2&=& 2 \, U_{\vec{\pi}^2},\;\;\;M_3^2 = 2 \, U_{\vec{\pi}^2}, \el
M_4^2&=&  \frac{1}{2} \left[ 2\, U_{\vec{\pi}^2} + 4 \, \vec{\pi}^2 \,
  U_{\vec{\pi}^2 \vec{\pi}^2} + U_{\sigma \sigma} -
  \sqrt{(2\, U_{\vec{\pi}^2} + 4\, \vec{\pi}^2
      U_{\vec{\pi}^2 \vec{\pi}^2} - U_{\sigma \sigma} )^2 
          + 16\, \vec{\pi}^2\, U_{\sigma \vec{\pi}^2}^2
   } \right]. 
\ee
For $\vec{\pi}^2=0$, the masses of the three pion modes 
are degenerate. 
The symmetry breaking terms in the $\sigma$-direction do not affect the
$O(3)$-symmetry of the pion subspace, so that the pion
fields appear only in the combination $\vec{\pi}^2$ in the
eigenvalues. The constituent quark mass is given by
\be
M_q^2&=&g^2 [(\sigma + m_c)^2 +\vec{\pi}^2]
\label{eq:mqexpansion}
\ee
To derive renormalization group flow equations 
in a finite four-dimensional volume $L^3 \times T$,
we replace the integrals over the momenta in the evaluation of the trace
(\ref{eq:1loop}) by a sum
\begin{equation}
\int
dp_{i}\,\ldots\rightarrow\frac{2\pi}{L}\sum_{n_i=-\infty}^{\infty}\ldots\,
\end{equation}
We are free in the choice of boundary conditions for the bosons and fermions in
the space directions. However, in the time direction, the boundary
conditions are fixed by the
statistics of the fields. Adopting the language of lattice literature,
we use $T$ to denote  the  
length of the finite volume box in Euclidean time direction . Then the
Matsubara frequencies take the values 
\be
\omega _{n_0} =\frac{2\pi n_0}{T}\quad\mathrm{and}\quad \nu _{n_0}=\frac{(2 n_0
+1)\pi}{T}\,. 
\ee
for bosons and for fermions, respectively. 
In the following we use the short-hand notation
\begin{equation}\label{eq:fv_mom}
p_{p}^{2}= \frac{4\pi^{2}}{L^{2}}
\sum_{i=1}^{3} n_{i}^{2}\quad\mathrm{and}\quad
p_{ap}^{2}=\frac{4\pi^{2}}{L^{2}} \sum_{i=1}^{3}\left( 
n_{i}+\frac{1}{2}\right)^{2}
\end{equation}
for the three-momenta in the case of periodic (p) and anti-periodic
(ap) boundary 
conditions. The flow equation corresponding to
eq.~(\ref{eq:IV_fe}) is
\be
k\frac{\partial}{\partial
  k} U_{k}(\sigma,\vec{\pi}^2,T,L)=&{\displaystyle
\frac{k^{2(a+1)}}{TL^{3}}\sum_{n_0} \sum_{\vec{n}}\Bigg(-\frac{4N_{c}
  N_{f}}{(k^{2}+\nu 
_{n_0} ^2 +p_{ap,p}^{2}+M_{q}^{2}(\sigma,\vec{\pi}^{2}))^{a+1}}} \nn \\ &
{\displaystyle \sum_{i=1}^{N_f^2=4}
\frac{1}{(k^{2}+\omega _{n_0} ^2
+p_{p}^{2}+M_{i}^{2}(\sigma,\vec{\pi}^{2}))^{a+1}} \Bigg)}.\,\label{eq:FV_fe}
\ee
The sums in eq. (\ref{eq:FV_fe}) run from $-\infty$ to $+\infty$, where the
vector $\vec{n}$ denotes
$(n_{1},n_{2},n_{3})$. For both finite and infinite volume, we choose
$a=2$ for the cutoff function. For a volume with infinite extent in time 
direction $T\rightarrow\infty$, we perform the sum over the
Matsubara frequencies analytically \cite{Braun:2003ii}:
\be
k\frac{\partial}{\partial
  k} U_{k}(\sigma,\vec{\pi}^2,
T\rightarrow\infty,L)=&{\displaystyle\frac{3}{16}\frac{k^{6}}{L^{3}}
  \sum_{\vec{n  }}\Bigg(-\frac{4
  N_{c} N_{f}}{(k^{2}+p_{ap,p}^{2}+M_{q}^{2}(\sigma,\vec{\pi}^{2}))^{5/2}} }\nn
\\ & {\displaystyle \sum_{i=1}^{N_f^2=4}
\frac{1}{(k^{2}+p_{p}^{2}+M_{i}^{2}(\sigma,\vec{\pi}^{2}))^{5/2}}\Bigg)}\,
\label{eq:FV_t0}  
\ee
Note that we employ both flow equations (\ref{eq:FV_fe}) and
(\ref{eq:FV_t0}) for 
our numerical calculations in the next section. We would like to make
one comment: 
The insertion of the regulator function (\ref{eq:cutoff-fct}) in the
Schwinger proper-time integral is not necessary to regularize the infrared
regime, since finite volume calculations are already infrared
finite. However, if we keep the volume fixed, the insertion of the
regulator function   
is needed to integrate out the quantum fluctuations in a controlled way.  

In order to solve the partial differential equations (\ref{eq:IV_fe}) and
(\ref{eq:FV_fe}), we project these flow equations on the following ansatz for
the mesonic potential \cite{Braun:2004yk}:
\be
U_{k}(\sigma, \vec{\pi}^2)&=& \sum_{i=0}^{N_\sigma} \sum_{j=0}^{i+j
 \le N_\sigma} a_{ij}(k) (\sigma-\sigma_0(k))^i (\sigma^2+\vec{\pi}^2
-\sigma_0(k)^2)^j  \label{eq:pot_ansatz}.
\ee
Performing such a projection, we get, in principle, an infinite set of coupled
first-order differential equations. To solve this set of equations, we have to
truncate the ansatz at some power in $(\sigma-\sigma_0(k))$ and
$(\sigma^2+\vec{\pi}^2
-\sigma_0(k)^2)$. In this
paper, we use $N_\sigma =2$. The resulting finite set of flow
equations can be solved straightforwardly in a numerical calculation.

\section{Calculation}
\label{sec:calculation}

We have solved the RG flow equations numerically and
will present the results for the volume dependence of
the pion mass and the pion decay constant in the following section. First  
we discuss some details about the numerical evaluation and the determination
of 
the coefficients of the ansatz for the potential eq. \eqref{eq:pot_ansatz} at
the UV scale.
At the ultraviolet cutoff scale $\Lambda_{UV}$, the meson potential can be
characterized by 
the values of the couplings $m_{UV}$ and $\lambda _{UV}$ in eq.
\eqref{eq:pot_UV}. All other coefficients in the ansatz eq.
\eqref{eq:pot_ansatz} are set to zero. In order to solve the flow equations for
the effective potential, we truncate the ansatz for the potential as
discussed in the last section. In the present work, we expand the potential
up to mass dimension four in the
mesonic fields. Furthermore, we have to specify a value for the
current quark mass 
$m_c$, which controls the degree of explicit symmetry breaking. The Yukawa
coupling $g$ does not evolve in the present approximation
\cite{Jungnickel:1995fp, Berges:1997eu, Schaefer:1999em}. 
We choose $g=3.26$, which leads to a reasonable constituent quark mass 
of $M_q = g (f_\pi +m_c) \approx 310\;\mathrm{MeV}$ for physical values for the
pion decay constant
$f_\pi=93 \;\mathrm{MeV}$ and the current quark mass $g m_c=7\;\mathrm{MeV}$. 

\begin{table}
\begin{ruledtabular}
\begin{tabular}{rrrrrr}
$\Lambda _{UV}\;\; \mathrm{[MeV]}$ & $m_{UV}\;\; \mathrm{[MeV]}$ &
  $\lambda_{UV}$ & $g m_c \; \mathrm{[MeV]}$ &  $f_\pi \;\;\mathrm{[MeV]}$ &
$m_\pi \;
  \mathrm{[MeV]}$ \\
\hline
1500 & 779.0 & 60 &  2.10 &  90.38 & 100.08\\
1500 & 747.7 & 60 &  9.85 &  96.91 & 200.1\phantom{0}\\
1500 & 698.0 & 60 & 25.70 & 105.30 & 300.2\phantom{0}\\
\end{tabular}
\end{ruledtabular}
\caption{\label{tab:start} Values for the parameters 
  at the $UV$-scale used in the numerical evaluation. We determine
  these parameters by fitting in infinite volume to a particular pion
  mass $m_\pi(\infty)$ and the 
  corresponding value of the pion decay constant $f_\pi(\infty)$,
  taken from chiral 
  perturbation theory. In our notation, the physical current
  quark mass corresponds to $g m_c$. We use $g=3.26$.}
\end{table}

In table~\ref{tab:start}, we summarize the three parameter sets
which we use in obtaining our results for pion masses of
$100$, $200$ and $300\;\mathrm{MeV}$, see also
ref.~\cite{Braun:2004yk}. In our comparison of 
different boundary conditions, we use the same
parameter sets to obtain results for either periodic or anti-periodic
boundary conditions for the fermions.
We determine these UV parameters by fitting to 
the pion mass $m_\pi(\infty)$ and to the corresponding pion decay
constant $f_\pi(\infty)$,  
which is taken in 
infinite volume from chiral perturbation theory \cite{Colangelo:2003hf}. We
then evolve the RG equations with these parameters to predict the volume
dependence of $f_\pi(L)$ and $m_\pi(L)$. 
From table~\ref{tab:start}, we can read off that the pion mass is
primarily determined by the value of the 
current quark mass, which controls the explicit symmetry breaking. To obtain
the correct value for the pion decay constant for a given pion mass,
the meson mass at the UV scale $m_{UV}$
has to be
decreased from $780\; \mbox{MeV}$ to $700\;\mbox{MeV}$ when the  pion
mass increases from $100$ 
to $300\; \mathrm{MeV}$. We use the same value for the
four-meson-coupling $\lambda_{UV}$ 
for all values of the pion mass and pion decay constant
considered here.
The possible values of the current quark mass are limited by the
requirement that all masses, in particular the sigma-mass, must remain
substantially smaller than the  ultraviolet cutoff $\Lambda_{UV} \approx 1500
\;\mathrm{MeV}$ of the
model. We have checked that our results are to a large
degree independent of the particular choice of UV parameters: different
sets of starting parameters give the same volume dependence, provided
that they lead to the same values of the pion mass and
pion decay constant in infinite volume.

We use the result of chiral perturbation theory for the dependence of
the pion decay constant
on the pion mass to facilitate the comparison between the quark meson
model and chPT. However, 
it is possible to get the correct behavior of the pion decay constant as a
function of a single symmetry breaking
parameter with renormalization group methods, as was shown in
infinite volume 
\cite{Jungnickel:1997yu}. 

For anti-periodic boundary conditions for the fermions, we have
previously investigated the dependence of our results on the cutoff scale
$\Lambda_{UV}$ \cite{Braun:2004yk} and found it to be small for
light pions, and only moderate for the largest pion mass we
considered here. We argued that such a trend towards a stronger
cutoff dependence for larger pion masses was to be
expected, since the existence of a cutoff becomes more relevant for
heavier mesons. This analysis still pertains to the results with
anti-periodic boundary conditions presented here. For a choice of
periodic boundary conditions, however, the cutoff dependence of the
results is somewhat more pronounced, in particular for small volumes
when the Euclidean time extent is kept large.  Varying
the cutoff between $1.5\;\mbox{GeV}$ and  $1.1\;\mbox{GeV}$ for a
pion mass of $m_\pi(\infty)=300 \; \mbox{MeV}$, we find that the largest
variations are of the order of $5 - 6 \%$ of the pion mass, and take place in
a volume range of $L=0.5 - 1.0 \; \mbox{fm}$, depending on the exact
ratio $T/L$ of time and space extent. As we argue below, this is
mainly due to effects on the quark condensation: for periodic boundary
conditions, a larger UV cutoff allows for the build-up of a larger
condensate in finite volume, since for any given volume, a larger
number of momentum
modes $ 2 \pi |\vec{n}|  / L $ remain below the cutoff and
contribute. In a volume region where the quarks dominate the finite
volume effects, a certain cutoff dependence of these effects is
therefore expected in this model.

The sums over the momentum modes in the flow equations cannot be performed
analytically, therefore we have to truncate the sums at a maximal mode number
$N_{max}=\mathrm{max}|\vec{n}|$. With this truncation, 
we introduce an additional UV cutoff in our calculation. In order to guarantee
that this cutoff does not affect our results we require
\be
\frac{2\pi}{L} N_{max} \gg \Lambda _{UV}\,.
\ee
We have to take care that this relation is well satisfied since we are
using a "soft" cutoff 
function. We have checked the dependence of the results on $N_{max}$
in \cite{Braun:2004yk} and found that
it is sufficient to use $N_{max}=40$ for $\Lambda
_{UV}=1.5\,\mathrm{GeV}$ and volumes up to $L=5 \; \mbox{fm}$, which
we will also use for the calculations in this paper.
The numerical evaluation of the sums over the
momentum modes simplifies significantly if we take the box sides as integer
multiples of some length scale $L_0$, such that $L= n_L L_0$ and $T=
n_T L_0$. Therefore we restrict ourselves to this case. Below,  
we show that the results for the low-energy observables strongly 
depend on the ratio $T/L = n_T/n_L$.

\section{Results}
\label{sec:results}

We have calculated the pion mass shift
\be
R[m_{\pi}(L)] = \frac{m_{\pi}(L)-m_{\pi}(\infty)}{m_{\pi}(\infty)}
\ee
with both choices for the fermionic boundary conditions for three
different pion  masses, $m_\pi(\infty)=100,\,200$ and $300\,\mathrm{MeV}$,
and for infinite ($T/L \to \infty$) as well as for finite extent of the
Euclidean time axis with different ratios $T/L=3/1,\,3/2,\,1/1$.

\begin{figure}
\includegraphics[scale=0.80, clip=true, angle=0,
  draft=false]{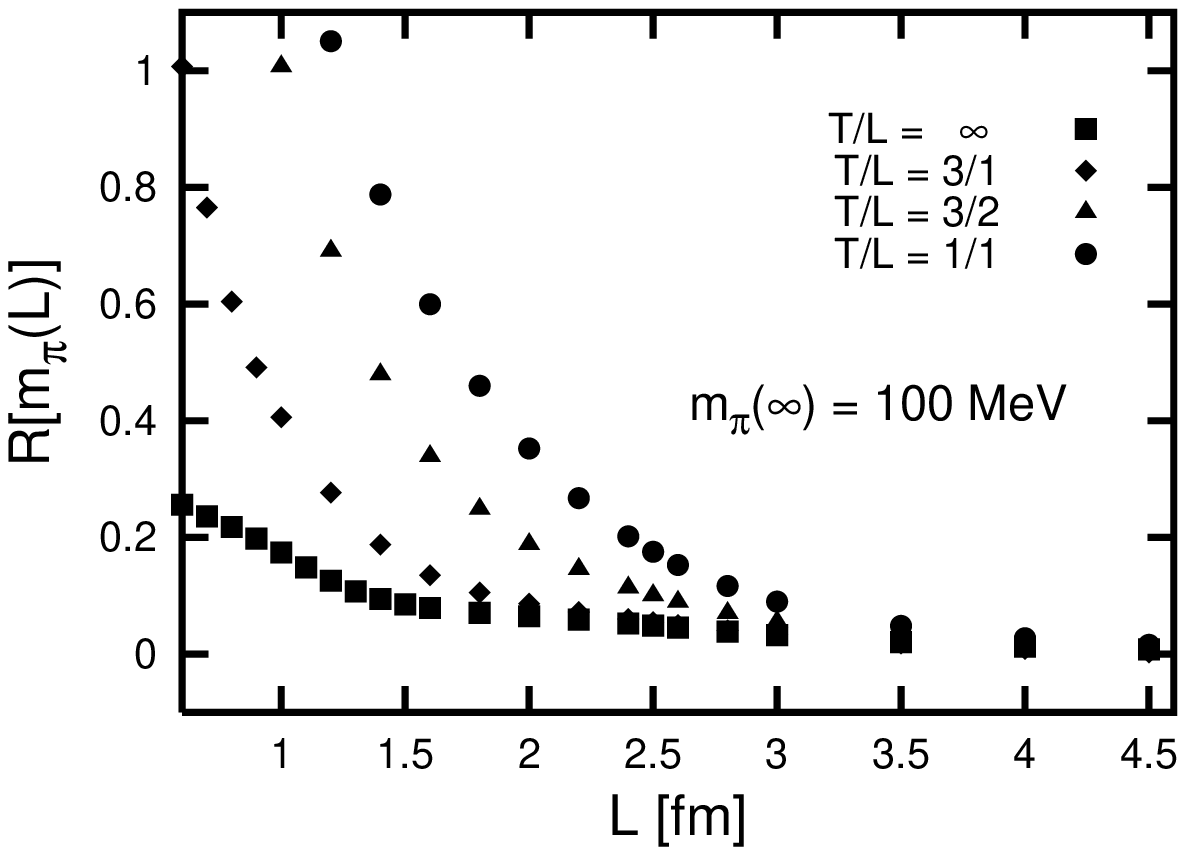} 
\includegraphics[scale=0.80, clip=true, angle=0,
  draft=false]{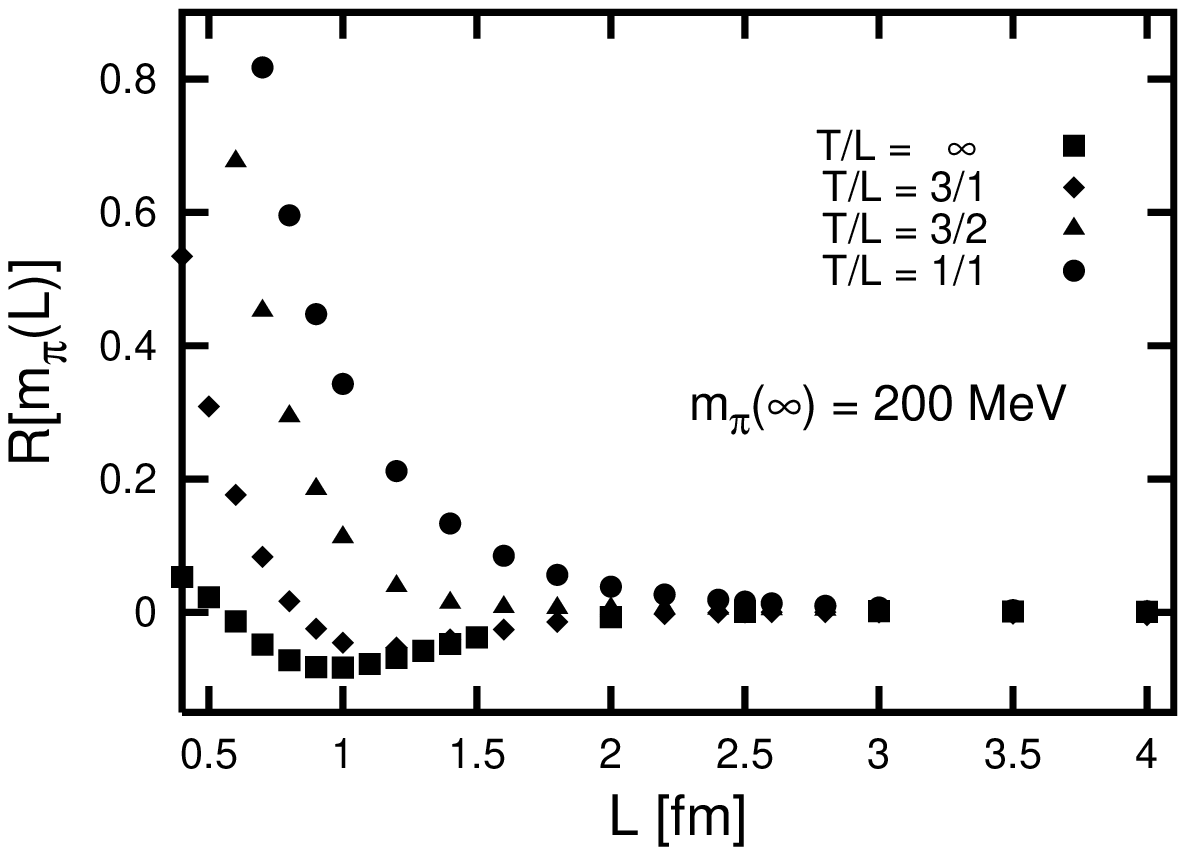} 
\includegraphics[scale=0.80, clip=true, angle=0,
  draft=false]{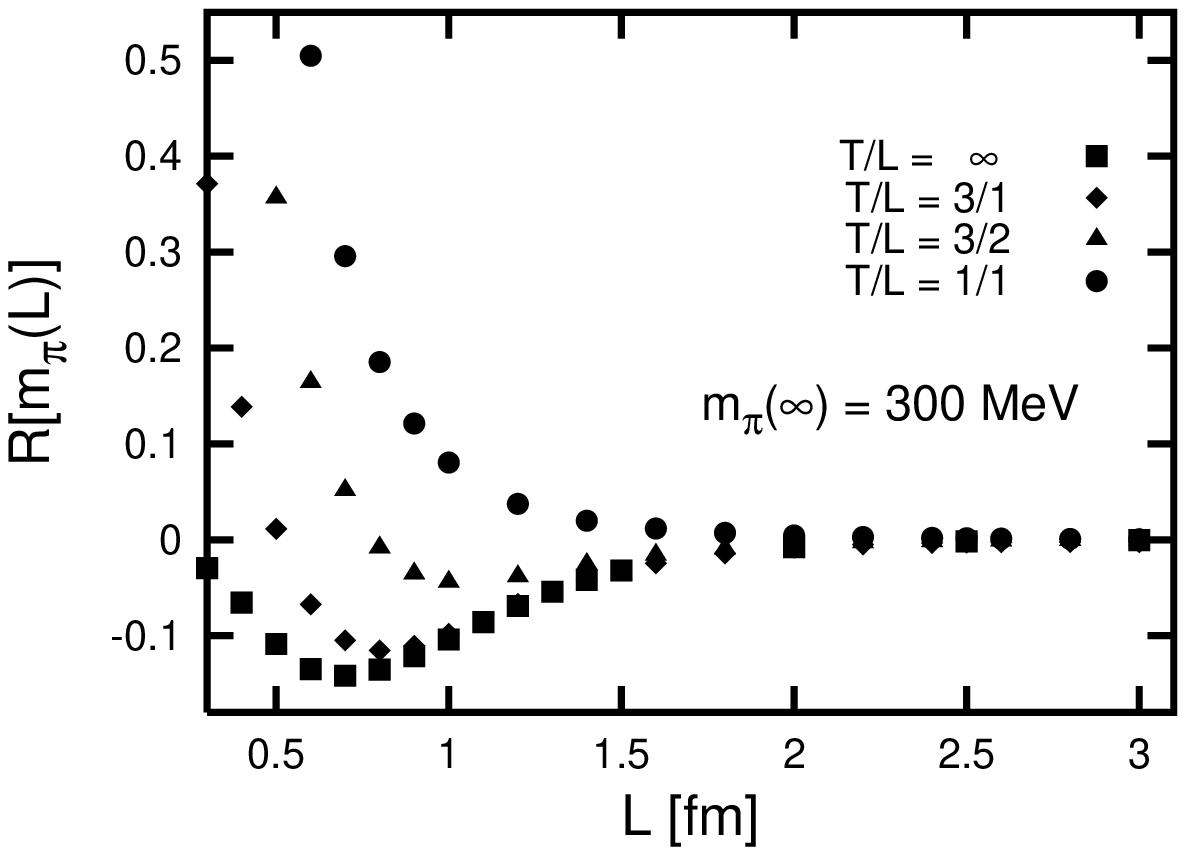}
\caption{\label{fig:pure_pb} Results for the pion mass shift $R[m_\pi(L)] =
(m_\pi(L)-m_\pi(\infty))/m_\pi(\infty)$, in a finite Euclidean volume
  of size $V=L^3 \times T$, for periodic boundary conditions. 
  The ratio of $T/L$ for the different curves is
  given in the figure. We show the results for pion masses of
  $m_\pi(\infty) =100, 200, 300 \; \mbox{MeV}$ (identified in the figure).}
\end{figure}

In Fig. \ref{fig:pure_pb}, we show the results for the pion mass shift
with periodic boundary conditions as a function of the box size
$L$. The three panels show the results for the three different pion
masses we investigated, and the curves are labeled with the ratios
$T/L$. 
The main new and surprising observation is that in this case, for
certain volume ranges, 
the mass of the pion in the finite volume can be {\it lower} than in
infinite volume.  
In particular, this is the case  for pion masses $m_{\pi} (\infty)\geq
 200\,\mathrm{MeV}$, ratios $T/L \geq 3/2$, and volumes smaller than
$2\,\mathrm{fm}$: $R[m_\pi(L)]$ takes on negative values and develops
a minimum. This can be seen in the lower two panels of Fig.
\ref{fig:pure_pb}. Secondly, we note that this minimum in the mass
shift becomes deeper for larger pion masses $m_{\pi} (\infty)$, and
the corresponding larger values of $f_\pi(\infty)$. For $m_{\pi}
(\infty)=300\,\mathrm{MeV}$, the  
pion mass shift reaches down to approximately $R[m_\pi]=-0.14$ at
$L=0.7\,\mathrm{fm}$. 

\begin{figure}
\includegraphics[scale=0.80, clip=true, angle=0,
  draft=false]{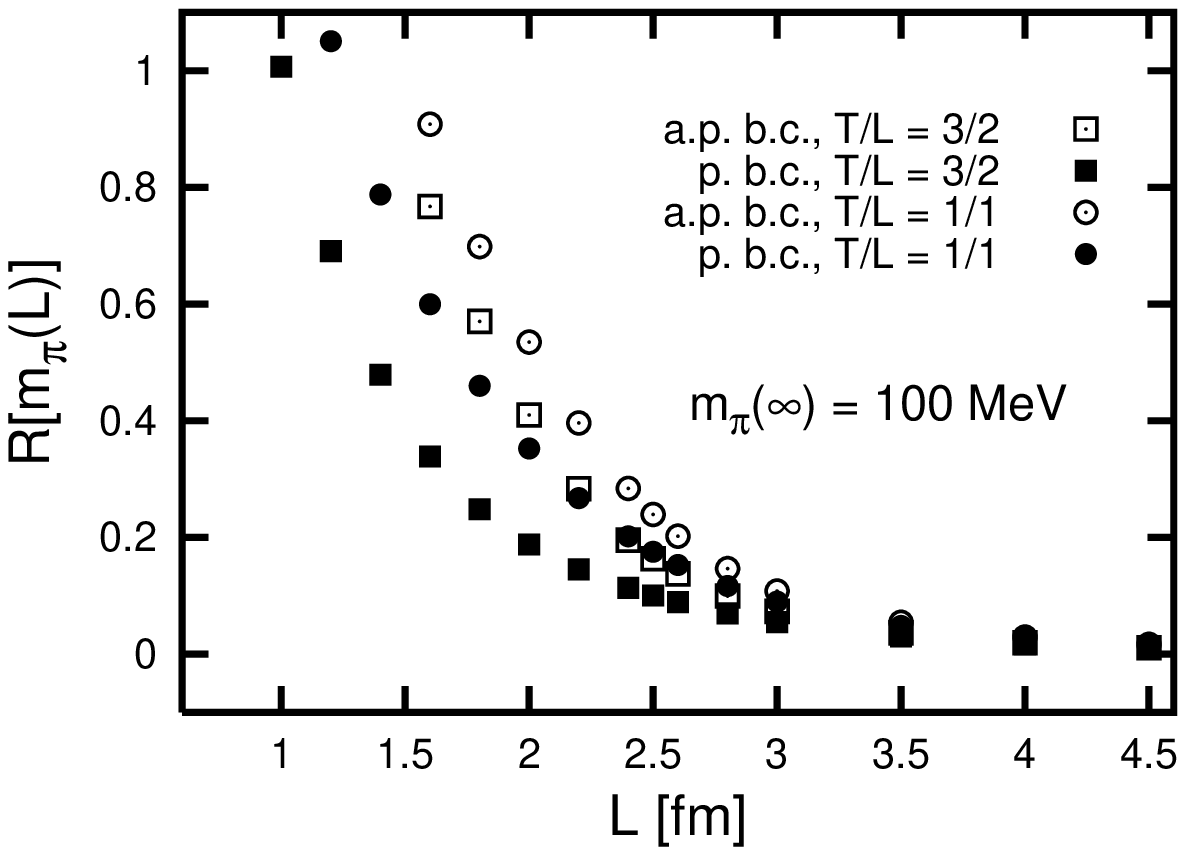}
\includegraphics[scale=0.80, clip=true, angle=0,
  draft=false]{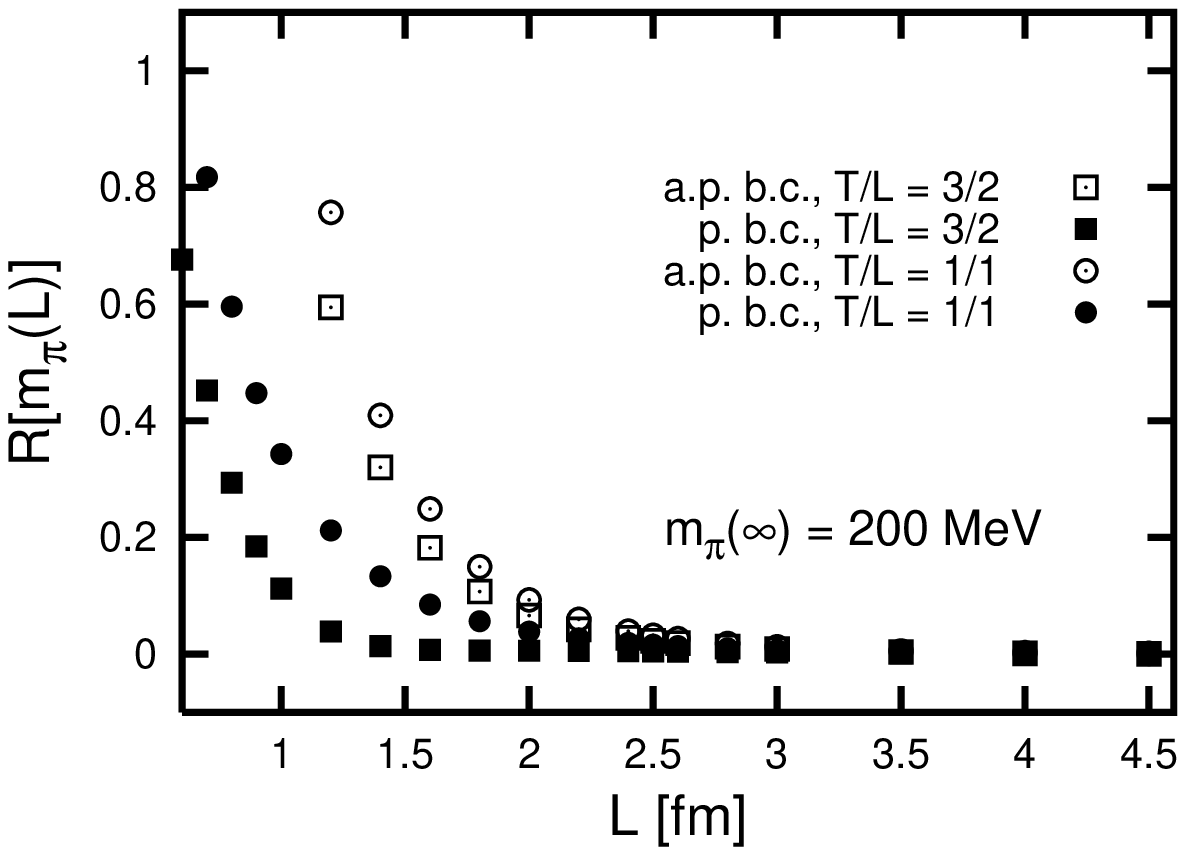} 
\includegraphics[scale=0.80, clip=true, angle=0,
  draft=false]{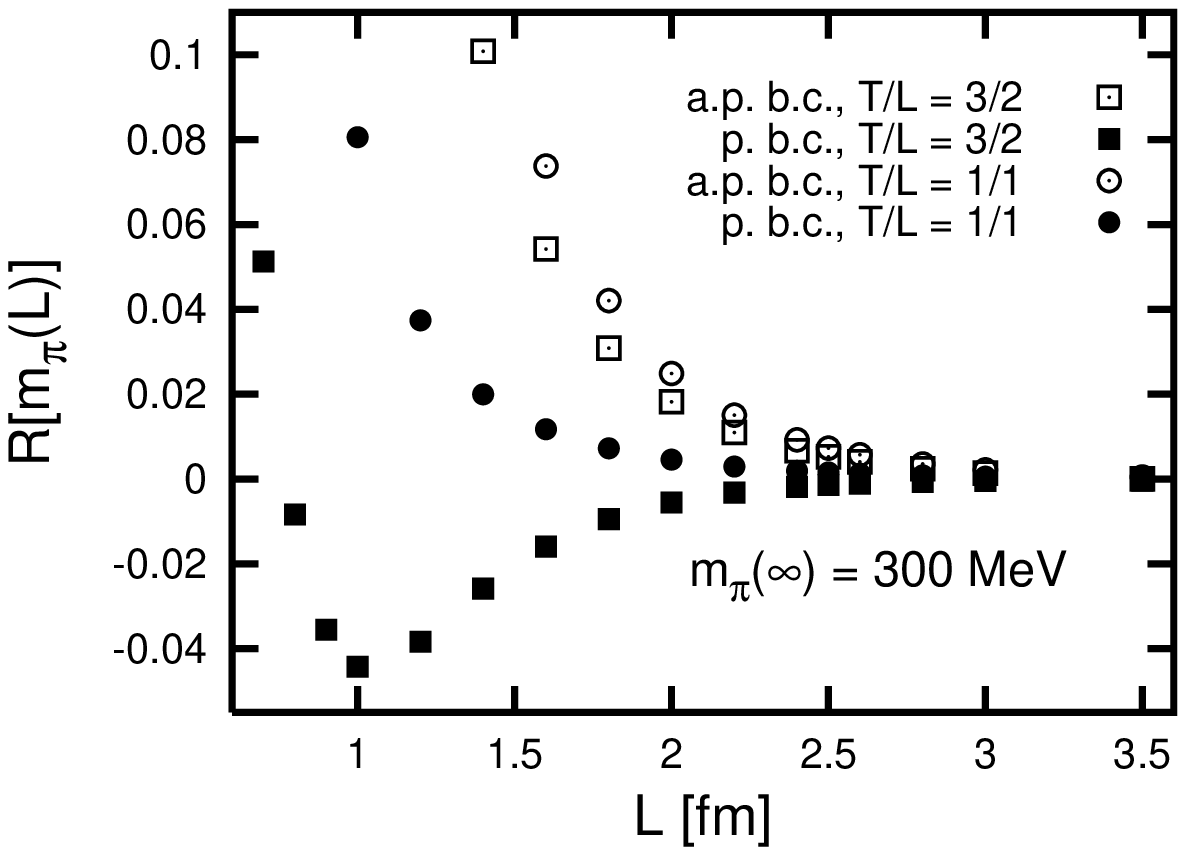}
\caption{\label{fig:compbc} Comparison of the pion mass shift in
  finite volume $R[m_\pi(L)] =
(m_\pi(L)-m_\pi(\infty))/m_\pi(\infty)$ for the two choices of fermionic
   boundary conditions. Open symbols denote results for anti-periodic,
  solid symbols for periodic boundary condition. The size of the
  volume is $V=L^3 \times T$, the ratios of $T/L$ for
  the different curves are
  given in the figures. We show results for pion masses of
  $m_\pi(\infty) =100, 200, 300 \; \mbox{MeV}$ (identified in the figure).}
\end{figure}

In Fig. \ref{fig:compbc}, we
compare the results for the pion mass shift with periodic (p. b.c.)
and anti-periodic (a.p. b.c.) 
boundary conditions for the fermion
fields, for the ratios $T/L = 3/2$ and  $T/L = 1/1$. 
Clearly, employing p. b.c. lowers the relative mass shift $R[m_{\pi}(L)]$,
compared to using a.p. b.c.. The differences become larger in
smaller volumes, for larger pion masses $m_{\pi} (\infty)$, and with
increasing ratios $T/L$. 
As we have seen, if the length of the box in the
Euclidean time direction is taken to infinity, for large pion masses
the pion mass can be smaller in finite than in infinite volume, so that 
the finite volume shift becomes negative.

Although at first a surprising result, this shift to smaller pion
masses can actually be explained in the framework of the quark-meson
model and its mechanism of chiral symmetry breaking. In order to show
this, we resort to a version of the model that is simplified
compared to our ansatz \eqref{eq:pot_ansatz}, but still contains the
same essential structure. In this model, for a fixed symmetry breaking
parameter $g m_c$, the pion mass is completely specified by the scale-dependent
order parameter $\sigma_0(k, L)$, and by the values given at the UV
scale for the coupling $g$ and the meson mass $m_{UV}^2$. According to
\cite{Walecka:1995mi, Zinn-Justin:2002ru}, it is 
\be\label{eq:mpi_fpi}
M_\pi^2(k,L) = \frac{m_c m_{UV} ^2}{ \sigma_0(k,L)}.
\ee
For periodic boundary conditions, the ``squeezing'' of the quark fields
in a small finite volume leads to an increase in the chiral quark
condensate, before a further decrease of
the volume size induces a restoration of chiral symmetry.
Following eq.~\eqref{eq:mpi_fpi}, the increase in the order parameter
leads in turn to the observed decrease in the pion mass. 

The intermediate increase in the order parameter with the decreasing
volume size 
can be explained more rigorously from the flow equations. Since this
increase occurs in volumes that are already quite small, 
the flow is dominated by 
the zero-momentum modes and it is sufficient to analyze the contributions
of these modes. 

The zero mode contribution to the flow equation from quarks and mesons 
is for purely periodic boundary conditions in spatial directions given by
\be
\label{eq:U0}
\left[k \frac{\partial}{\partial
  k} U_k(\sigma, \vec{\pi}^2, L, T)
  \right]_0&=&-\frac{k^{2(a+1)}}{T L^3} \bigg(2 \cdot \frac{4N_c
  N_f}{(k^2 + \nu_0^2 + 
  M_q(k,L, \sigma, \vec{\pi}^2 ) ^2)^{(a+1)}} \nn \\
 &&\hspace{-1.5cm}-\frac{N_f ^2 -1}{(k^2 + M_{\pi}(k,L, \sigma,
    \vec{\pi}^2 ) ^2)^{(a+1)}}-\frac{1}{(k^2 + M_{\sigma}(k,L, \sigma,
    \vec{\pi}^2 ) ^2)^{(a+1)}}\bigg)
\ee
where $\nu_0^2 = (\pm \pi/T)^2$ corresponds to the value of the two 
Matsubara frequencies closest to zero. The prefactor $1/L^3$ diverges 
for $L \to 0$ for all momentum modes, but enhances only  
the zero modes: For the non-zero momentum modes, the enhancement is
canceled and they are in fact strongly suppressed, which is due to
the factors $1/L^2$ of the momentum terms in the denominators. 
If we scale $T$ proportional to $L$, because of the Matsubara
frequencies this suppression occurs also
for the lowest fermionic terms, although it is much weaker. 
The result of this competition between suppression and 
enhancement for the fermions depends on the ratio $T/L$.

We first consider exclusively the contributions of the fermionic zero
modes, which exist only for periodic boundary conditions:
\be
\left[k\frac{\partial}{\partial
  k} U_{k}(\sigma,\vec{\pi}^2,T,L)\right]^F_0&=&
-\frac{k^{2(a+1)}}{TL^{3}}\cdot  2 \cdot \frac{4N_{c} N_{f}}{(k^{2}+\nu _{0} ^2
+M_{q}^{2}(\sigma,\vec{\pi}^{2}))^{a+1}}\,
\label{eq:pot_fermion}
\ee
This truncation to the fermionic contributions only is equivalent 
to the leading term of a large $N_c$-approximation, as it is shown in
\cite{Meyer:2001zp}. In principle, eq.~\eqref{eq:pot_fermion}
can be integrated analytically, since the constituent quark mass,
given by $M_q^2(\sigma, \vec{\pi}^2)=g^2[(\sigma+m_c)^2 +
  \vec{\pi}^2]$, does not depend on any scale-dependent quantities.
The result shows that the zero mode contributions to the potential as a
function of the expectation value are {\it repulsive} for small
values. Consequently, these contributions increase the expectation
value $\sigma_0(k,  L)$ 
and thus the value of the pion decay constant. Since these zero-momentum
contributions are enhanced for small volumes, this explains the
increase in the expectation value.

Alternatively, this can be understood in more detail by a direct analysis of
the zero mode contributions to the flow equation for the minimum
$\sigma_0(k, L)$ of the potential.  
Since 
the flow equation for $\sigma_0(k, L)$ is obtained from the minimum condition 
\be
&&\frac{\partial}{\partial \sigma} U_k(\sigma = \sigma_0(k, L),
\vec{\pi}^2 =0, L, T) = 0,
\label{eq:minimum}
\ee
it is determined by the flow of the potential. As we have seen in our
analysis above, the fermionic contributions
tend to increase the absolute value of the minimum $\sigma_0(k, L)$,
while the mesonic contributions tend to 
decrease it. 
Thus, we can perform this analysis entirely by considering the
zero mode part of the potential flow given in eq.~\eqref{eq:U0}.

The renormalization scale $k$ controls the momenta of the quantum
fluctuations that are integrated out. As soon as this momentum scale
drops below the mass of one of the degrees of freedom, that particular field
can no longer contribute to the RG evolution of the running couplings:
it decouples from the RG flow. We restrict the discussion here to
scales $k<m_{\sigma}$, where the sigma meson has already decoupled.

With periodic boundary conditions, the finite box length in the
Euclidean time direction $T$ is the only scale which affects the zero
modes. The scale $\pi/T$ is in competition with the renormalization
scale $k$, and if $k$ drops below this scale, 
the lowest Matsubara frequency $\nu_0=\pi/T$ acts as a cutoff and stops that 
part of the evolution which is driven by the quark fields. 
If $T$ is sufficiently small, this happens already above the scale at
which chiral symmetry breaking sets in. In that case, condensation of
the quark fields is prevented, and the constituent quark mass remains
small. This means that $m_\pi(k \to 0, L)$ remains large and that
$R[m_\pi(L)]$ is large and positive. This is illustrated by the
results for $T/L=1/1$ and small $L$ in Fig.~\ref{fig:pure_pb}. 

The situation is different for large values of $T/L>3/2$. Here, the
additional scale set by $1/T$ plays a less important role and becomes
relevant only for much smaller volumes.
In this case, quarks build up a large condensate. According to 
eq. \eqref{eq:mpi_fpi}, 
this increase in the chiral condensate leads to a decrease of the pion mass,
which is visible in Fig.~\ref{fig:pure_pb} for $T/L>3/2$,
$m_\pi(\infty)\geq200\,\mathrm{MeV}$, and $L\geq 0.8\,\mathrm{fm}$. 
For large values of $T/L$, the decrease in the condensate for small
volumes cannot be explained by the presence of the cutoff $\pi/T$ for
the quark fields alone. There is an additional mechanism that
decreases $\sigma_0$ in such a way that chiral symmetry is broken less
strongly. 
For very small volumes, the pion contributions in eq.~\eqref{eq:U0}
dominate the flow of $\sigma_0$. 
Even for a large ratio $T/L$, this leads to a decrease in $\sigma_0$
and the observed rise in $R[m_\pi(L)]$ for small $L$. 

For anti-periodic boundary conditions, we do not find any decrease of
$R[m_\pi(L)]$ with decreasing finite volume size $L$ for any value of $T/L$, 
as can be seen in the comparison in  Fig.~\ref{fig:compbc}. 
As we have argued in \cite{Braun:2004yk}, in our RG approach with anti-periodic
boundary conditions,
two effects are responsible for the finite volume behavior: effects
due to the quark condensation, and effects due to light pions which appear 
after the chiral condensate has been built up by the quark fields. 
In contrast to the case of 
periodic boundary conditions, for anti-periodic boundary conditions
the formation of the quark 
condensate is strongly suppressed by the lowest 
possible momentum for the fermions, which  
is $\sqrt{3}\pi /L$, see eq.~\eqref{eq:fv_mom}, and acts
as an infrared cutoff. 
Consequently, for small $L$, fewer modes contribute to the
chiral condensate. If in addition $T/L$ is small, the condensate
decreases further and we observe a larger mass shift  $R[m_\pi(L)]$.

\begin{figure}
\includegraphics[scale=0.80, clip=true, angle=0,
  draft=false]{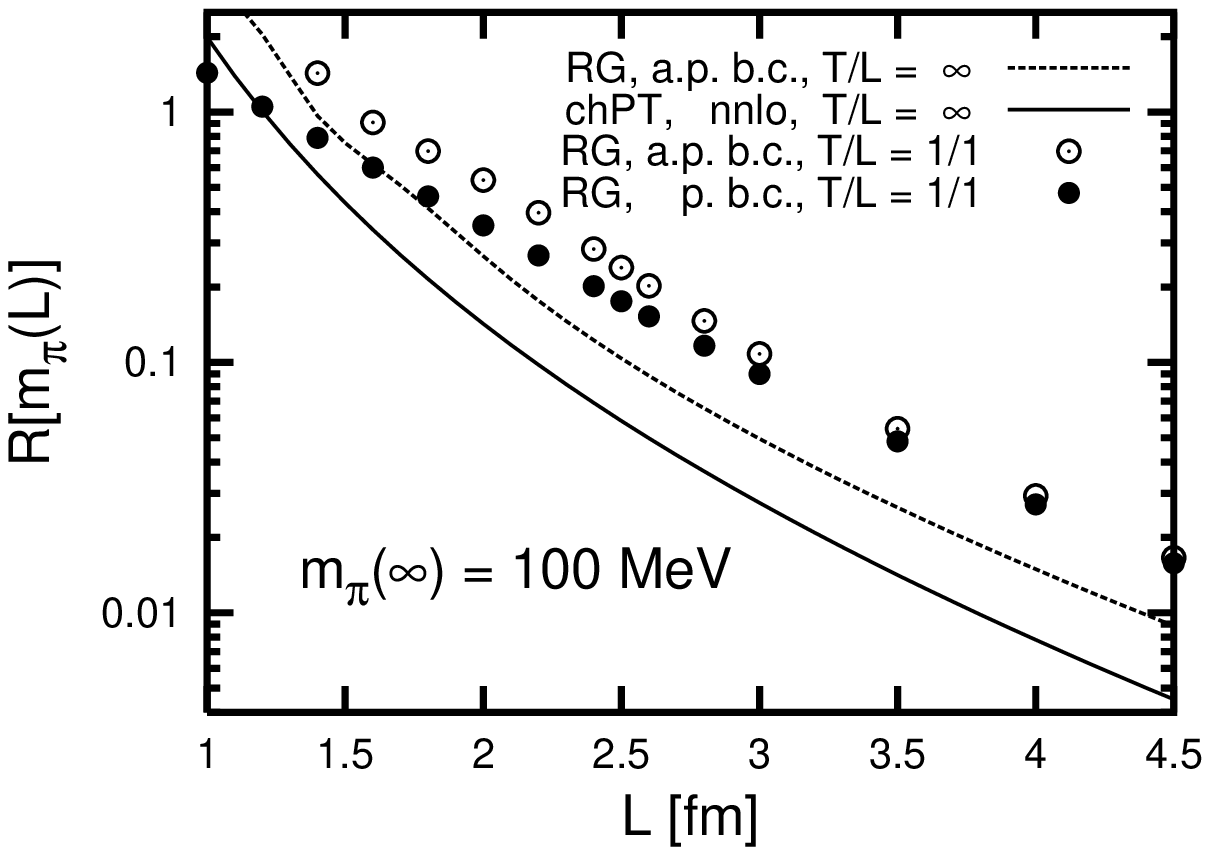}
\includegraphics[scale=0.80, clip=true, angle=0,
  draft=false]{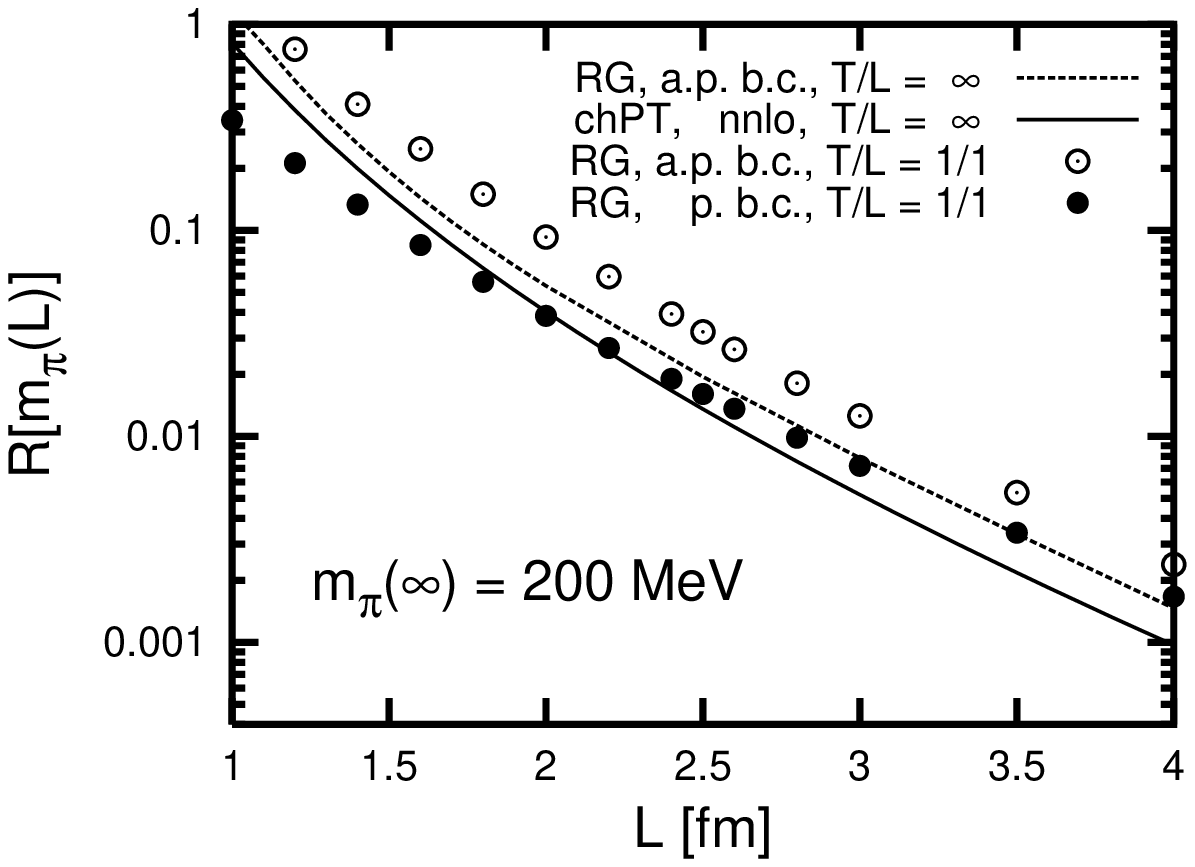}
\includegraphics[scale=0.80, clip=true, angle=0,
  draft=false]{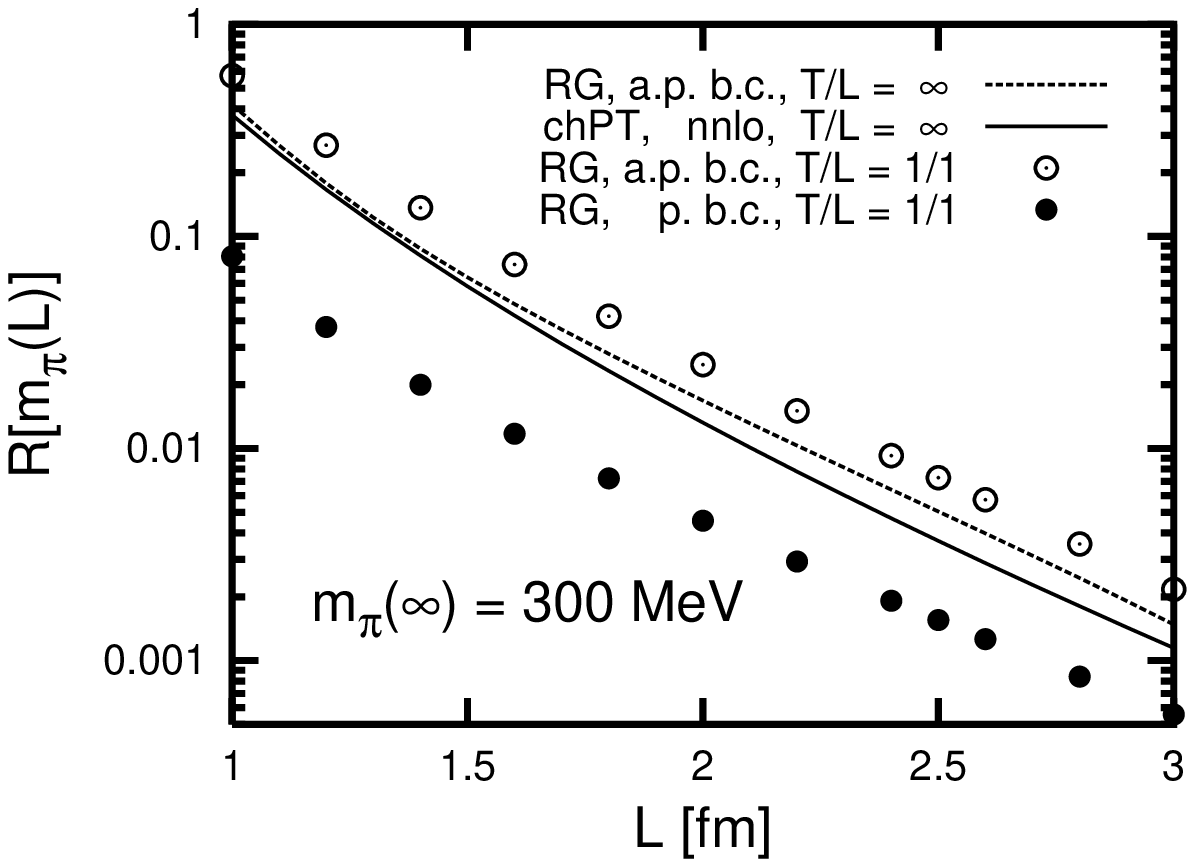}
\caption{\label{fig:chPT} Comparison of the pion mass shift $R[m_\pi(L)] =
(m_\pi(L)-m_\pi(\infty))/m_\pi(\infty)$ for
  different boundary conditions with the results of chiral perturbation theory
\cite{Colangelo:2005gd}
  on a logarithmic scale.  The ratio of $T/L$ for the different curves is
  given in the figures. We show results for a pion mass of
  $m_\pi(\infty) =100, 200, 300 \; \mbox{MeV}$ (identified in the figure).}
\end{figure}

Finally, in Fig.~\ref{fig:chPT} we compare our results for the pion
mass shift to the results of chiral perturbation theory. (Note that
Fig.~\ref{fig:chPT} has a
logarithmic scale, whereas Figs.~\ref{fig:pure_pb} and
\ref{fig:compbc} have linear scales.) 
We present results for different pion masses from RG calculations with
both periodic and anti-periodic 
boundary conditions for the fermions, and from chiral perturbation
theory \cite{Colangelo:2003hf, Colangelo:2005gd}. For the chPT
results, the pion mass shift is calculated with the help of
L\"uscher's formula \cite{Luscher:1985dn}, which relates the leading
corrections of the pion mass in finite Euclidean volume to the
$\pi\pi$-scattering amplitude in infinite volume. The sub-leading
corrections drop as $\mathcal{O}(e^{-\bar{m}L})$ with
$\bar{m}\geq\sqrt{3/2}\, m_\pi$. Using a
calculation of the $\pi\pi$-scattering amplitude in chPT to three
loops ({\it nnlo}) as input
for L\"uscher's formula, the authors of ref.~\cite{Colangelo:2003hf} obtain a 
correction above the leading order, which is then added to the
one-loop result of Gasser and Leutwyler \cite{Gasser:1987ah}. 
L\"uscher's original approach only considers the
  periodicity of pion propagators in finite volume as an invariance
  under a shift by $L$. More recently, in ref.~\cite{Colangelo:2005gd}
  this has been improved to account for the fact that these propagators
  are actually invariant under shifts by $\vec{n} L$ with arbitrary $\vec{n}$.
  The result is a L\"uscher formula resummed
  over $\vec{n}$, which is very similar to the original one. The finite volume
  shift for the pion mass is significantly increased by this resummation.
In \cite{Braun:2004yk} we have carefully compared our RG results with
anti-periodic boundary conditions to the chPT results from
\cite{Colangelo:2003hf}. Here, we use the improved results from
\cite{Colangelo:2005gd} for the comparison. 
The RG results are still consistently above the 
results from chPT. L\"uscher's approach becomes an increasingly
better approximation with increasing pion mass for a given volume
size. The decreasing differences between the chPT results
and the RG results with increasing pion mass
are compatible with this estimate. For large
volumes, the mass shift is completely controlled by pion effects and
drops as $e^{-m_\pi L}$, so that both the RG and the chPT results have the
same slope in the logarithmic plot. 
For the entire volume range shown in
Fig.~\ref{fig:chPT}, the RG and chPT results apparently differ only
by a factor 
which is almost independent of the volume size. For $m_\pi(\infty) =
300 \; \mathrm{MeV}$, the chPT and RG results agree within errors.
For small volumes, however, the RG
approach has the advantage that it can be extended to describe the 
transition into a regime with approximately restored chiral symmetry, where
the chiral expansion becomes unreliable.

The mesonic degrees of freedom are less affected
by the ratio $T/L$. The upper curve in
Fig.~\ref{fig:chPT} represents 
RG calculations with anti-periodic boundary conditions and $T/L=1/1$,
which gives a larger $R[m_\pi(L)]$ compared to the lower curve  
corresponding to $T/L=\infty$.
Fluctuations due to the light pions yield a decrease of the
condensate and explain the increase of $R[m_\pi(L)]$ for larger volumes.
In particular for small pion masses ($m_\pi=100 \; \mbox{MeV}$) and
the ratio $T/L=1/1$, the results with periodic and with anti-periodic
boundary conditions overlap over a wide volume range. From our analysis, for
sufficiently small values of $m_{\pi} (\infty)$ this
is expected in the region where pion dynamics dominate. Because of
this, the slopes of the curves are very similar.
The deviations between results at the same, fixed ratio $T/L$ that differ
only in the choice of boundary 
conditions become larger for increasing pion masses $m_{\pi} (\infty)$
and decay constants $f_{\pi} (\infty)$.
This indicates that fermionic effects are increasingly
important. Evidence for this is also the observation that
the results 
for the pion mass shift with periodic boundary conditions have a
smaller slope, compared to the results with
anti-periodic boundary conditions, and also compared to those of
chPT. The reason 
is that the cutoff scales are different: for periodic boundary
conditions, the lowest fermion momentum 
mode is given by the lowest Matsubara frequency $\nu_0=\pi/T$, and not
determined by $\sqrt{3}\pi/L$ as for anti-periodic boundary
conditions.   
In particular for large values of $T/L$, this explains that the finite
volume mass
shift will be much larger for anti-periodic
boundary conditions. For small
volumes, we thus find the importance of quark effects confirmed by the
dependence on the boundary conditions. But since pion effects dominate
for larger volumes, the results of chPT and of our RG approach
converge for large $L$.

\section{Comparison to lattice results and conclusions}
\label{sec:conclusion}

Apart from the general interest of finite volume effects, the main
  motivation for our  
  current investigation is its possible 
application to lattice gauge theory.
At present, most lattice calculations are performed in volumes of
  the order of $L= 2-3 \; \mathrm{fm}$. 
In recent systematic studies of finite
  volume effects done with Wilson fermions, the lightest pion
  masses are of the order of $m_\pi = 400 - 500 \; \mathrm{MeV}$
  \cite{AliKhan:2001tx, AliKhan:2003cu, Allton:2001sk,
  Aoki:2002uc,Guagnelli:2004ww, Orth:2005kq}.
With staggered fermions, pion masses as low as $250 \;
  \mathrm{MeV}$ have been realized \cite{Aubin:2004fs}.
Simulations with fermions with good chiral properties such as
  domain-wall or overlap fermions have been done with pion
  masses as low as $180 \; \mathrm{MeV}$ in the quenched approximation
  \cite{Aoki:2002vt, Chen:2003im} and as low as $360\; \mathrm{MeV}$ with two
  fully dynamical flavors \cite{Aoki:2004ht}.
Because the finite volume effects depend on the mass of the lightest
  field, they become more severe for smaller pion masses.
Thus, the better the statistical accuracy of these calculations, the
more important it becomes to understand finite size effects and to
control the finite size extrapolation. 

Our model incorporates chiral symmetry and can still be used in the
vicinity of the point where chiral symmetry is restored. 
Finite volume effects should therefore be
captured as far as they relate to chiral symmetry breaking.
But it is not a gauge theory, there are no gluons, and consequently the
constituent quarks in this model are not confined. There is no
guarantee that the same mechanisms apply as in QCD. 
Since the
  model contains dynamical meson fields and chiral symmetry is broken
  in the usual way, our results can only be compared directly to those
  of unquenched lattice calculations with two dynamical quark flavors,
  where normal chPT is also applicable.
However, qualitatively our arguments regarding the quark
  condensate may also
  have implications for quenched simulations, since a similar mechanism
  may apply.

Extrapolations to infinite volume using chiral perturbation theory are
extremely successful in the description of 
the volume dependence of nucleon properties, such as for example the
nucleon mass \cite{AliKhan:2003cu, Bedaque:2004dt}. 
However, as far as meson masses are concerned, the finite volume mass
shifts observed on the lattice deviate from the predictions of chiral
perturbation theory. This holds also \cite{Aoki:2002uc} for
L\"uscher's approach \cite{Luscher:1985dn}, which only takes
pion effects into account as well. 
Generally, the predicted mass shifts are
much smaller than the observed ones \cite{Aoki:1993gi, Aoki:2002uc,
  Guagnelli:2004ww, Orth:2005kq}. The inclusion of higher orders in
the chiral expansion \cite{Colangelo:2003hf} and a summation of
additional contributions in L\"uscher's expression
\cite{Colangelo:2005gd} increase the size of the predicted mass
shifts and decrease the distance to the RG results.
For the physical values of the pion mass and the pion
  decay constant, chiral perturbation theory can be applied for
  volume sizes $L \gg 1 \mathrm{fm}$, but a priori it is impossible to
  say how large exactly the volume has to be, according to
  ref.~\cite{Colangelo:2005gd}. Ultimately, this question can only be 
  answered by lattice calculations.

Comparing our results to those from chiral perturbation theory, we find
agreement for larger pion masses, provided we
impose anti-periodic boundary conditions on the fermionic fields. 
As expected, the differences increase for very small volumes, where
chiral symmetry restoration becomes important. For periodic boundary
conditions, large pion masses, and a large ratio $T/L$, the mass shifts
in small volumes behave differently from those of chPT. 
In particular, even in volume sizes as large as $L=2.5 \;
\mathrm{fm}$ and for $T \to \infty$, the results for the relative
shift of the pion mass can as much as double under a change of
boundary conditions, for example from $R[m_\pi(L)]=0.0488$ with
periodic to $R[m_\pi(L)]=0.1037$ with 
anti-periodic boundary conditions for a pion mass of $m_\pi = 100 \;
\mathrm{MeV}$. In Table~\ref{tab:bounds} we give bounds on the minimum
size of the volume that are necessary to keep the finite volume mass
shift smaller than $10\%$ resp. $1\%$, calculated for periodic and
antiperiodic boundary conditions in the RG approach, and for
comparison from the NNLO
chPT calculation of ref.~\cite{Colangelo:2005gd}. In general, periodic boundary
  conditions allow to achieve the same accuracy with regard to finite
  volume effects with smaller volume sizes than anti-periodic boundary
  conditions.
\begin{table}
\begin{ruledtabular}
\begin{tabular}{llll}
$m_\pi(\infty)$  &  & 
$R[m_\pi(L)] < 0.1$ & $R[m_\pi(L)] < 0.01$ \\
\hline
$100 \; \mathrm{MeV}$ & RG, ap. & $L > 2.523 \; \mathrm{fm}$ & $L > 4.381 \; \mathrm{fm}$ \\
     & RG, p.  & $L > 1.351 \; \mathrm{fm}$ & $L > 4.259 \; \mathrm{fm}$ \\
     & chPT    & $L > 2.187 \; \mathrm{fm}$ & $L > 3.785 \; \mathrm{fm}$ \\ 
\hline
$ 200 \; \mathrm{MeV}$ & RG, ap. & $L > 1.736 \; \mathrm{fm}$ & $L > 2.842 \; \mathrm{fm}$ \\
     & RG, p.  & $L > 0.5\phantom{00} \; \mathrm{fm}^{*}$& $L > 1.888 \; \mathrm{fm}$ \\
     & chPT    & $L > 1.639 \; \mathrm{fm}$ & $L > 2.653 \; \mathrm{fm}$ \\
\hline
$300 \; \mathrm{MeV}$  & RG, ap. & $L > 1.359 \; \mathrm{fm}$ & $L > 2.213 \; \mathrm{fm}$ \\
     & RG, p.  & $L > 1.022 \; \mathrm{fm}$ & $L > 1.911 \; \mathrm{fm}$ \\
     & chPT    & $L > 1.339 \; \mathrm{fm}$ & $L > 2.104 \; \mathrm{fm}$ \\
\end{tabular}
\end{ruledtabular}
\caption{\label{tab:bounds} Bounds on the minimum size of the volume
  $V=L^3 \times T$ for $T \to \infty$ such that the finite volume pion mass
  shift $R[m_\pi(L)]$ is $< 0.1$ or $< 0.01$, for different values
  of the pion mass $m_\pi(\infty)$. RG results are given for
  anti-periodic and for periodic boundary conditions, chPT results are
  those in NNLO obtained in \cite{Colangelo:2005gd}. Note that for
  periodic boundary conditions and for $m_\pi=200\;  \mathrm{MeV}$ and
  $ m_\pi=  300 \; \mathrm{MeV}$, the bounds are set by a decrease
  of the pion mass. ${}^{*}$For $m_\pi=200\; \mathrm{MeV}$ and
  periodic boundary conditions, the bound $R[m_\pi(L)]< 0.1$ is satisfied in
  the full volume range described by our model 
  (cf. also Fig.~\ref{fig:pure_pb}).}  
\end{table}

For periodic boundary conditions, our results reproduce the
  qualitative behavior of the lattice results, but clearly differ from
  chPT. For anti-periodic boundary conditions, they largely agree with chPT.
This suggests that for anti-periodic boundary
  conditions, the effective low-energy constants relevant for the
  finite volume effects of our observables agree with those of chPT,
  in agreement with the argument by Gasser 
  and Leutwyler for QCD \cite{Gasser:1987zq}. However, the differences
  then might imply that the low-energy constants change for
  periodic boundary conditions.

The issue of finite volume effects has been addressed in several
lattice studies 
\cite{Fukugita:1992wq, Aoki:1993gi, Aoki:2002uc, Guagnelli:2004ww,
  Orth:2005kq}. 
The pion mass shift $R[m_{\pi}(L)]$ calculated by the ZeRo
collaboration \cite{Guagnelli:2004ww}, which is shown in
Fig.~\ref{fig:zero}, actually becomes negative and has a minimum at
small volume sizes. Although this negative shift is small, the
result seems to be significant.
The minimum is most pronounced for small quark masses  (at $\kappa=0.1350$).
The position of the minimum corresponds to $m_{\pi} L=3.5$ or
$L=1.264\,\mathrm{fm}$ with $T/L=2.25$. The results were obtained in
the quenched approximation with periodic boundary conditions for the
quark fields.  
Similar observations have also been made in 
\cite{Aoki:2002uc, Orth:2005kq}, where the simulations were performed
with dynamical Wilson quarks.  

In our calculation, such a decrease in the pion mass is reproduced if
we choose periodic boundary conditions for the quarks. The minimum
appears for large pion 
mass $m_{\pi} (\infty)=300\,\mathrm{MeV},\,T/L\geq3/2$ and
$L=1\,\mathrm{fm}$, cf. Fig.~\ref{fig:pure_pb}. Our model suggests a
mechanism for the appearance of this minimum, which may be the same
mechanism as on the lattice.
In contrast to our findings, however, the decrease of the pion
mass in finite volume seems to be larger for smaller infinite-volume
pion mass. 
For lattice calculations, several other mechanisms for
finite volume mass shifts have been suggested, from an interaction of
hadrons with their mirror states on a periodic 
lattice \cite{Fukugita:1992wq} to effects on quark propagation
related to a breaking of the center 
symmetry of the gauge group \cite{Aoki:1993gi}.

The influence of boundary conditions for sea and valence quarks in
lattice simulations was also studied by Aoki {\it et al.}
\cite{Aoki:1993gi}. They find that periodic boundary conditions lead  
to a lower mass shift than anti-periodic boundary conditions (see table
III of \cite{Aoki:1993gi}). This finding is in agreement with our
results, as can be seen in Figs.~\ref{fig:compbc} and \ref{fig:chPT}. 
The actual pion mass on the lattice is very high ($>1\,\mathrm{GeV}$).
Different choices for the boundary conditions of sea and valence
quarks make it possible for the
authors to establish a connection between the mass shift and the
expectation value that Polyakov loops acquire in the presence of sea
quarks. They relate the large increase of the pion mass observed for
small lattice size to the restoration of chiral symmetry. This is
illustrated by their results for the chiral condensate (Fig.~10 of
\cite{Aoki:1993gi}), which decreases strongly in small volumes.
In the same figure, the condensate may increase
for intermediate volume size, which would be 
similar to the behavior of the order parameter seen in our simple model.
We agree that the mass shift in small volumes is due to chiral
symmetry restoration, and reproduce this result in our calculations.

Our RG approach improves our understanding of the mechanisms of
finite volume effects in QCD,  
but cannot yet give a model independent extrapolation formula to
relate finite lattice results to the hadronic world. 

In conclusion, we have discussed effects of quark boundary
conditions on the finite volume shifts of the pion mass. We used the
framework of an RG treatment of the quark-meson model to offer
a possible mechanism which accounts for quark effects. 
Our approach shows the importance of the 
fermionic boundary conditions for the pion mass and the 
pion decay constant. The differences between the results for
periodic and anti-periodic
boundary conditions increase for increasing pion mass and increasing
ratio $T/L$.
Our analysis agrees qualitatively with the
observations from lattice QCD, in regards to the dependence on quark boundary
conditions as well as in regards to an apparent drop of the pion mass in finite
volume. We find convergence of our results to those of chiral
perturbation theory calculations for large pion masses and large
volumes, where quark effects are not important.

\acknowledgments
The authors would like to thank G. Colangelo for providing the data
for the comparison to chPT in Fig.~\ref{fig:chPT}. B.K. would like
to thank I. Wetzorke and K. Jansen for a useful discussion. J.B. would
like to thank the GSI for financial support. This work is supported in
part by the Helmholtz association under grant no. VH-VI-041, and in
part by the EU Integrated Infrastructure Initiative Hadron Physics
(I3HP) under contract RII3-CT-2004-506078.  

\bibliography{rgfvbc}

\begin{thebibliography}{39}
\expandafter\ifx\csname natexlab\endcsname\relax\def\natexlab#1{#1}\fi
\expandafter\ifx\csname bibnamefont\endcsname\relax
  \def\bibnamefont#1{#1}\fi
\expandafter\ifx\csname bibfnamefont\endcsname\relax
  \def\bibfnamefont#1{#1}\fi
\expandafter\ifx\csname citenamefont\endcsname\relax
  \def\citenamefont#1{#1}\fi
\expandafter\ifx\csname url\endcsname\relax
  \def\url#1{\texttt{#1}}\fi
\expandafter\ifx\csname urlprefix\endcsname\relax\def\urlprefix{URL }\fi
\providecommand{\bibinfo}[2]{#2}
\providecommand{\eprint}[2][]{\url{#2}}

\bibitem[{\citenamefont{Procura et~al.}(2004)\citenamefont{Procura, Hemmert,
  and Weise}}]{Procura:2003ig}
\bibinfo{author}{\bibfnamefont{M.}~\bibnamefont{Procura}},
  \bibinfo{author}{\bibfnamefont{T.~R.} \bibnamefont{Hemmert}},
  \bibnamefont{and} \bibinfo{author}{\bibfnamefont{W.}~\bibnamefont{Weise}},
  \bibinfo{journal}{Phys. Rev.} \textbf{\bibinfo{volume}{D69}},
  \bibinfo{pages}{034505} (\bibinfo{year}{2004}), \eprint{hep-lat/0309020}.

\bibitem[{\citenamefont{Arndt and Lin}(2004)}]{Arndt:2004bg}
\bibinfo{author}{\bibfnamefont{D.}~\bibnamefont{Arndt}} \bibnamefont{and}
  \bibinfo{author}{\bibfnamefont{C.~J.~D.} \bibnamefont{Lin}},
  \bibinfo{journal}{Phys. Rev.} \textbf{\bibinfo{volume}{D70}},
  \bibinfo{pages}{014503} (\bibinfo{year}{2004}), \eprint{hep-lat/0403012}.

\bibitem[{\citenamefont{Ali~Khan et~al.}(2004)}]{AliKhan:2003cu}
\bibinfo{author}{\bibfnamefont{A.}~\bibnamefont{Ali~Khan}} \bibnamefont{et~al.}
  (\bibinfo{collaboration}{QCDSF-UKQCD}), \bibinfo{journal}{Nucl. Phys.}
  \textbf{\bibinfo{volume}{B689}}, \bibinfo{pages}{175} (\bibinfo{year}{2004}),
  \eprint{hep-lat/0312030}.

\bibitem[{\citenamefont{Colangelo and D{\"u}rr}(2004)}]{Colangelo:2003hf}
\bibinfo{author}{\bibfnamefont{G.}~\bibnamefont{Colangelo}} \bibnamefont{and}
  \bibinfo{author}{\bibfnamefont{S.}~\bibnamefont{D{\"u}rr}},
  \bibinfo{journal}{Eur. Phys. J.} \textbf{\bibinfo{volume}{C33}},
  \bibinfo{pages}{543} (\bibinfo{year}{2004}), \eprint{hep-lat/0311023}.

\bibitem[{\citenamefont{Colangelo and Haefeli}(2004)}]{Colangelo:2004xr}
\bibinfo{author}{\bibfnamefont{G.}~\bibnamefont{Colangelo}} \bibnamefont{and}
  \bibinfo{author}{\bibfnamefont{C.}~\bibnamefont{Haefeli}},
  \bibinfo{journal}{Phys. Lett.} \textbf{\bibinfo{volume}{B590}},
  \bibinfo{pages}{258} (\bibinfo{year}{2004}), \eprint{hep-lat/0403025}.

\bibitem[{\citenamefont{Colangelo et~al.}(2005)\citenamefont{Colangelo,
  D{\"u}rr, and Haefeli}}]{Colangelo:2005gd}
\bibinfo{author}{\bibfnamefont{G.}~\bibnamefont{Colangelo}},
  \bibinfo{author}{\bibfnamefont{S.}~\bibnamefont{D{\"u}rr}}, \bibnamefont{and}
  \bibinfo{author}{\bibfnamefont{C.}~\bibnamefont{Haefeli}}
  (\bibinfo{year}{2005}), \eprint{hep-lat/0503014}.

\bibitem[{\citenamefont{Bedaque et~al.}(2005)\citenamefont{Bedaque,
  Griesshammer, and Rupak}}]{Bedaque:2004dt}
\bibinfo{author}{\bibfnamefont{P.~F.} \bibnamefont{Bedaque}},
  \bibinfo{author}{\bibfnamefont{H.~W.} \bibnamefont{Griesshammer}},
  \bibnamefont{and} \bibinfo{author}{\bibfnamefont{G.}~\bibnamefont{Rupak}},
  \bibinfo{journal}{Phys. Rev.} \textbf{\bibinfo{volume}{D71}},
  \bibinfo{pages}{054015} (\bibinfo{year}{2005}), \eprint{hep-lat/0407009}.

\bibitem[{\citenamefont{Gasser and Leutwyler}(1988)}]{Gasser:1987zq}
\bibinfo{author}{\bibfnamefont{J.}~\bibnamefont{Gasser}} \bibnamefont{and}
  \bibinfo{author}{\bibfnamefont{H.}~\bibnamefont{Leutwyler}},
  \bibinfo{journal}{Nucl. Phys.} \textbf{\bibinfo{volume}{B307}},
  \bibinfo{pages}{763} (\bibinfo{year}{1988}).

\bibitem[{\citenamefont{Gasser and
  Leutwyler}(1987{\natexlab{a}})}]{Gasser:1986vb}
\bibinfo{author}{\bibfnamefont{J.}~\bibnamefont{Gasser}} \bibnamefont{and}
  \bibinfo{author}{\bibfnamefont{H.}~\bibnamefont{Leutwyler}},
  \bibinfo{journal}{Phys. Lett.} \textbf{\bibinfo{volume}{B184}},
  \bibinfo{pages}{83} (\bibinfo{year}{1987}{\natexlab{a}}).

\bibitem[{\citenamefont{Bernard et~al.}(2004)\citenamefont{Bernard, Hemmert,
  and Meissner}}]{Bernard:2003rp}
\bibinfo{author}{\bibfnamefont{V.}~\bibnamefont{Bernard}},
  \bibinfo{author}{\bibfnamefont{T.~R.} \bibnamefont{Hemmert}},
  \bibnamefont{and} \bibinfo{author}{\bibfnamefont{U.-G.}
  \bibnamefont{Meissner}}, \bibinfo{journal}{Nucl. Phys.}
  \textbf{\bibinfo{volume}{A732}}, \bibinfo{pages}{149} (\bibinfo{year}{2004}),
  \eprint{hep-ph/0307115}.

\bibitem[{\citenamefont{Bernard et~al.}(2005)\citenamefont{Bernard, Hemmert,
  and Meissner}}]{Bernard:2005fy}
\bibinfo{author}{\bibfnamefont{V.}~\bibnamefont{Bernard}},
  \bibinfo{author}{\bibfnamefont{T.~R.} \bibnamefont{Hemmert}},
  \bibnamefont{and} \bibinfo{author}{\bibfnamefont{U.-G.}
  \bibnamefont{Meissner}} (\bibinfo{year}{2005}), \eprint{hep-lat/0503022}.

\bibitem[{\citenamefont{Leinweber et~al.}(1999)\citenamefont{Leinweber, Lu, and
  Thomas}}]{Leinweber:1998ej}
\bibinfo{author}{\bibfnamefont{D.~B.} \bibnamefont{Leinweber}},
  \bibinfo{author}{\bibfnamefont{D.~H.} \bibnamefont{Lu}}, \bibnamefont{and}
  \bibinfo{author}{\bibfnamefont{A.~W.} \bibnamefont{Thomas}},
  \bibinfo{journal}{Phys. Rev.} \textbf{\bibinfo{volume}{D60}},
  \bibinfo{pages}{034014} (\bibinfo{year}{1999}), \eprint{hep-lat/9810005}.

\bibitem[{\citenamefont{Detmold et~al.}(2001)\citenamefont{Detmold,
  Melnitchouk, Negele, Renner, and Thomas}}]{Detmold:2001jb}
\bibinfo{author}{\bibfnamefont{W.}~\bibnamefont{Detmold}},
  \bibinfo{author}{\bibfnamefont{W.}~\bibnamefont{Melnitchouk}},
  \bibinfo{author}{\bibfnamefont{J.~W.} \bibnamefont{Negele}},
  \bibinfo{author}{\bibfnamefont{D.~B.} \bibnamefont{Renner}},
  \bibnamefont{and} \bibinfo{author}{\bibfnamefont{A.~W.}
  \bibnamefont{Thomas}}, \bibinfo{journal}{Phys. Rev. Lett.}
  \textbf{\bibinfo{volume}{87}}, \bibinfo{pages}{172001}
  (\bibinfo{year}{2001}), \eprint{hep-lat/0103006}.

\bibitem[{\citenamefont{Aoki et~al.}(2003)}]{Aoki:2002uc}
\bibinfo{author}{\bibfnamefont{S.}~\bibnamefont{Aoki}} \bibnamefont{et~al.}
  (\bibinfo{collaboration}{JLQCD}), \bibinfo{journal}{Phys. Rev.}
  \textbf{\bibinfo{volume}{D68}}, \bibinfo{pages}{054502}
  (\bibinfo{year}{2003}), \eprint{hep-lat/0212039}.

\bibitem[{\citenamefont{Guagnelli et~al.}(2004)}]{Guagnelli:2004ww}
\bibinfo{author}{\bibfnamefont{M.}~\bibnamefont{Guagnelli}}
  \bibnamefont{et~al.} (\bibinfo{collaboration}{Zeuthen-Rome (ZeRo)}),
  \bibinfo{journal}{Phys. Lett.} \textbf{\bibinfo{volume}{B597}},
  \bibinfo{pages}{216} (\bibinfo{year}{2004}), \eprint{hep-lat/0403009}.

\bibitem[{\citenamefont{Orth et~al.}(2005)\citenamefont{Orth, Lippert, and
  Schilling}}]{Orth:2005kq}
\bibinfo{author}{\bibfnamefont{B.}~\bibnamefont{Orth}},
  \bibinfo{author}{\bibfnamefont{T.}~\bibnamefont{Lippert}}, \bibnamefont{and}
  \bibinfo{author}{\bibfnamefont{K.}~\bibnamefont{Schilling}}
  (\bibinfo{year}{2005}), \eprint{hep-lat/0503016}.

\bibitem[{\citenamefont{Ali~Khan et~al.}(2002)}]{AliKhan:2001tx}
\bibinfo{author}{\bibfnamefont{A.}~\bibnamefont{Ali~Khan}} \bibnamefont{et~al.}
  (\bibinfo{collaboration}{CP-PACS}), \bibinfo{journal}{Phys. Rev.}
  \textbf{\bibinfo{volume}{D65}}, \bibinfo{pages}{054505}
  (\bibinfo{year}{2002}), \eprint{hep-lat/0105015}.

\bibitem[{\citenamefont{L{\"u}scher}(1986)}]{Luscher:1985dn}
\bibinfo{author}{\bibfnamefont{M.}~\bibnamefont{L{\"u}scher}},
  \bibinfo{journal}{Commun. Math. Phys.} \textbf{\bibinfo{volume}{104}},
  \bibinfo{pages}{177} (\bibinfo{year}{1986}).

\bibitem[{\citenamefont{Aoki et~al.}(1994{\natexlab{a}})}]{Aoki:1993gi}
\bibinfo{author}{\bibfnamefont{S.}~\bibnamefont{Aoki}} \bibnamefont{et~al.},
  \bibinfo{journal}{Phys. Rev.} \textbf{\bibinfo{volume}{D50}},
  \bibinfo{pages}{486} (\bibinfo{year}{1994}{\natexlab{a}}).

\bibitem[{\citenamefont{Aoki et~al.}(1994{\natexlab{b}})}]{Aoki:1993fq}
\bibinfo{author}{\bibfnamefont{S.}~\bibnamefont{Aoki}} \bibnamefont{et~al.},
  \bibinfo{journal}{Nucl. Phys. Proc. Suppl.} \textbf{\bibinfo{volume}{34}},
  \bibinfo{pages}{363} (\bibinfo{year}{1994}{\natexlab{b}}),
  \eprint{hep-lat/9311049}.

\bibitem[{\citenamefont{Sharpe}(1992)}]{Sharpe:1992ft}
\bibinfo{author}{\bibfnamefont{S.~R.} \bibnamefont{Sharpe}},
  \bibinfo{journal}{Phys. Rev.} \textbf{\bibinfo{volume}{D46}},
  \bibinfo{pages}{3146} (\bibinfo{year}{1992}), \eprint{hep-lat/9205020}.

\bibitem[{\citenamefont{Bernard and Golterman}(1992)}]{Bernard:1992mk}
\bibinfo{author}{\bibfnamefont{C.~W.} \bibnamefont{Bernard}} \bibnamefont{and}
  \bibinfo{author}{\bibfnamefont{M.~F.~L.} \bibnamefont{Golterman}},
  \bibinfo{journal}{Phys. Rev.} \textbf{\bibinfo{volume}{D46}},
  \bibinfo{pages}{853} (\bibinfo{year}{1992}), \eprint{hep-lat/9204007}.

\bibitem[{\citenamefont{Braun et~al.}(2005)\citenamefont{Braun, Klein, and
  Pirner}}]{Braun:2004yk}
\bibinfo{author}{\bibfnamefont{J.}~\bibnamefont{Braun}},
  \bibinfo{author}{\bibfnamefont{B.}~\bibnamefont{Klein}}, \bibnamefont{and}
  \bibinfo{author}{\bibfnamefont{H.~J.} \bibnamefont{Pirner}},
  \bibinfo{journal}{Phys. Rev.} \textbf{\bibinfo{volume}{D71}},
  \bibinfo{pages}{014032} (\bibinfo{year}{2005}), \eprint{hep-ph/0408116}.

\bibitem[{\citenamefont{Jungnickel and Wetterich}(1996)}]{Jungnickel:1995fp}
\bibinfo{author}{\bibfnamefont{D.~U.} \bibnamefont{Jungnickel}}
  \bibnamefont{and}
  \bibinfo{author}{\bibfnamefont{C.}~\bibnamefont{Wetterich}},
  \bibinfo{journal}{Phys. Rev.} \textbf{\bibinfo{volume}{D53}},
  \bibinfo{pages}{5142} (\bibinfo{year}{1996}), \eprint{hep-ph/9505267}.

\bibitem[{\citenamefont{Schaefer and Pirner}(1999)}]{Schaefer:1999em}
\bibinfo{author}{\bibfnamefont{B.-J.} \bibnamefont{Schaefer}} \bibnamefont{and}
  \bibinfo{author}{\bibfnamefont{H.-J.} \bibnamefont{Pirner}},
  \bibinfo{journal}{Nucl. Phys.} \textbf{\bibinfo{volume}{A660}},
  \bibinfo{pages}{439} (\bibinfo{year}{1999}), \eprint{nucl-th/9903003}.

\bibitem[{\citenamefont{Schaefer and Wambach}(2004)}]{Schaefer:2004en}
\bibinfo{author}{\bibfnamefont{B.-J.} \bibnamefont{Schaefer}} \bibnamefont{and}
  \bibinfo{author}{\bibfnamefont{J.}~\bibnamefont{Wambach}}
  (\bibinfo{year}{2004}), \eprint{nucl-th/0403039}.

\bibitem[{\citenamefont{Jungnickel and Wetterich}(1998)}]{Jungnickel:1997yu}
\bibinfo{author}{\bibfnamefont{D.~U.} \bibnamefont{Jungnickel}}
  \bibnamefont{and}
  \bibinfo{author}{\bibfnamefont{C.}~\bibnamefont{Wetterich}},
  \bibinfo{journal}{Eur. Phys. J.} \textbf{\bibinfo{volume}{C2}},
  \bibinfo{pages}{557} (\bibinfo{year}{1998}), \eprint{hep-ph/9704345}.

\bibitem[{\citenamefont{Braun et~al.}(2004)\citenamefont{Braun, Schwenzer, and
  Pirner}}]{Braun:2003ii}
\bibinfo{author}{\bibfnamefont{J.}~\bibnamefont{Braun}},
  \bibinfo{author}{\bibfnamefont{K.}~\bibnamefont{Schwenzer}},
  \bibnamefont{and} \bibinfo{author}{\bibfnamefont{H.-J.}
  \bibnamefont{Pirner}}, \bibinfo{journal}{Phys. Rev.}
  \textbf{\bibinfo{volume}{D70}}, \bibinfo{pages}{085016}
  (\bibinfo{year}{2004}), \eprint{hep-ph/0312277}.

\bibitem[{\citenamefont{Berges et~al.}(1999)\citenamefont{Berges, Jungnickel,
  and Wetterich}}]{Berges:1997eu}
\bibinfo{author}{\bibfnamefont{J.}~\bibnamefont{Berges}},
  \bibinfo{author}{\bibfnamefont{D.~U.} \bibnamefont{Jungnickel}},
  \bibnamefont{and}
  \bibinfo{author}{\bibfnamefont{C.}~\bibnamefont{Wetterich}},
  \bibinfo{journal}{Phys. Rev.} \textbf{\bibinfo{volume}{D59}},
  \bibinfo{pages}{034010} (\bibinfo{year}{1999}), \eprint{hep-ph/9705474}.

\bibitem[{\citenamefont{Walecka}(1995)}]{Walecka:1995mi}
\bibinfo{author}{\bibfnamefont{J.~D.} \bibnamefont{Walecka}},
  \bibinfo{journal}{Oxford Stud. Nucl. Phys.} \textbf{\bibinfo{volume}{16}},
  \bibinfo{pages}{1} (\bibinfo{year}{1995}).

\bibitem[{\citenamefont{Zinn-Justin}(2002)}]{Zinn-Justin:2002ru}
\bibinfo{author}{\bibfnamefont{J.}~\bibnamefont{Zinn-Justin}},
  \bibinfo{journal}{Int. Ser. Monogr. Phys.} \textbf{\bibinfo{volume}{113}},
  \bibinfo{pages}{1} (\bibinfo{year}{2002}).

\bibitem[{\citenamefont{Meyer et~al.}(2002)\citenamefont{Meyer, Schwenzer,
  Pirner, and Deandrea}}]{Meyer:2001zp}
\bibinfo{author}{\bibfnamefont{J.}~\bibnamefont{Meyer}},
  \bibinfo{author}{\bibfnamefont{K.}~\bibnamefont{Schwenzer}},
  \bibinfo{author}{\bibfnamefont{H.-J.} \bibnamefont{Pirner}},
  \bibnamefont{and} \bibinfo{author}{\bibfnamefont{A.}~\bibnamefont{Deandrea}},
  \bibinfo{journal}{Phys. Lett.} \textbf{\bibinfo{volume}{B526}},
  \bibinfo{pages}{79} (\bibinfo{year}{2002}), \eprint{hep-ph/0110279}.

\bibitem[{\citenamefont{Gasser and
  Leutwyler}(1987{\natexlab{b}})}]{Gasser:1987ah}
\bibinfo{author}{\bibfnamefont{J.}~\bibnamefont{Gasser}} \bibnamefont{and}
  \bibinfo{author}{\bibfnamefont{H.}~\bibnamefont{Leutwyler}},
  \bibinfo{journal}{Phys. Lett.} \textbf{\bibinfo{volume}{B188}},
  \bibinfo{pages}{477} (\bibinfo{year}{1987}{\natexlab{b}}).

\bibitem[{\citenamefont{Allton et~al.}(2002)}]{Allton:2001sk}
\bibinfo{author}{\bibfnamefont{C.~R.} \bibnamefont{Allton}}
  \bibnamefont{et~al.} (\bibinfo{collaboration}{UKQCD}),
  \bibinfo{journal}{Phys. Rev.} \textbf{\bibinfo{volume}{D65}},
  \bibinfo{pages}{054502} (\bibinfo{year}{2002}), \eprint{hep-lat/0107021}.

\bibitem[{\citenamefont{Aubin et~al.}(2004)}]{Aubin:2004fs}
\bibinfo{author}{\bibfnamefont{C.}~\bibnamefont{Aubin}} \bibnamefont{et~al.}
  (\bibinfo{collaboration}{MILC}), \bibinfo{journal}{Phys. Rev.}
  \textbf{\bibinfo{volume}{D70}}, \bibinfo{pages}{114501}
  (\bibinfo{year}{2004}), \eprint{hep-lat/0407028}.

\bibitem[{\citenamefont{Aoki et~al.}(2004{\natexlab{a}})}]{Aoki:2002vt}
\bibinfo{author}{\bibfnamefont{Y.}~\bibnamefont{Aoki}} \bibnamefont{et~al.},
  \bibinfo{journal}{Phys. Rev.} \textbf{\bibinfo{volume}{D69}},
  \bibinfo{pages}{074504} (\bibinfo{year}{2004}{\natexlab{a}}),
  \eprint{hep-lat/0211023}.

\bibitem[{\citenamefont{Chen et~al.}(2004)}]{Chen:2003im}
\bibinfo{author}{\bibfnamefont{Y.}~\bibnamefont{Chen}} \bibnamefont{et~al.},
  \bibinfo{journal}{Phys. Rev.} \textbf{\bibinfo{volume}{D70}},
  \bibinfo{pages}{034502} (\bibinfo{year}{2004}), \eprint{hep-lat/0304005}.

\bibitem[{\citenamefont{Aoki et~al.}(2004{\natexlab{b}})}]{Aoki:2004ht}
\bibinfo{author}{\bibfnamefont{Y.}~\bibnamefont{Aoki}} \bibnamefont{et~al.}
  (\bibinfo{year}{2004}{\natexlab{b}}), \eprint{hep-lat/0411006}.

\bibitem[{\citenamefont{Fukugita et~al.}(1993)\citenamefont{Fukugita, Mino,
  Okawa, Parisi, and Ukawa}}]{Fukugita:1992wq}
\bibinfo{author}{\bibfnamefont{M.}~\bibnamefont{Fukugita}},
  \bibinfo{author}{\bibfnamefont{H.}~\bibnamefont{Mino}},
  \bibinfo{author}{\bibfnamefont{M.}~\bibnamefont{Okawa}},
  \bibinfo{author}{\bibfnamefont{G.}~\bibnamefont{Parisi}}, \bibnamefont{and}
  \bibinfo{author}{\bibfnamefont{A.}~\bibnamefont{Ukawa}},
  \bibinfo{journal}{Nucl. Phys. Proc. Suppl.} \textbf{\bibinfo{volume}{30}},
  \bibinfo{pages}{365} (\bibinfo{year}{1993}).

\end{thebibliography}

\end{document}